\documentclass[onepage,12pt,a4paper,pdftex]{article}

\usepackage{graphicx, amsfonts, amsmath, amssymb, amscd,  theorem, exscale,
epic, eepic, epsfig}

\usepackage{fancyhdr,hyperref}

\usepackage{makeidx}
\makeindex

\newcounter{Figure}
\theoremstyle{plain}
\theoremheaderfont{\scshape}
\newtheorem{conj}{\bf Conjecture}
\newtheorem{Def}{\bf Definition}
\newtheorem{The}{\bf Theorem}

\newtheorem{prop}{\bf Proposition}
\theorembodyfont{\upshape}

\theoremheaderfont{\scshape\small} \theorembodyfont{\scshape\small}

\newcommand{\real}{ {\mathbb R} }

\newcommand{\slap}{\mbox{$ \triangle \mkern -13mu / \ $}}
\newcommand{\nlap}{\mbox{$ \nabla \mkern -13mu / \ $}}
\newcommand{\dlap}{\mbox{$ div \mkern -13mu / \ $}}
\newcommand{\Dlap}{\mbox{$ D \mkern -13mu / \ $}}

\newcommand{\clap}{\mbox{$ curl \mkern -23mu / \ $}}

\newcommand{\be}{\begin{equation}}
\newcommand{\ee}{\end{equation}}
\newcommand{\bea}{\begin{eqnarray}}
\newcommand{\eea}{\end{eqnarray}}
\newcommand{\beas}{\begin{eqnarray*}}
\newcommand{\eeas}{\end{eqnarray*}}

\newcommand{\Lie}{ {\mathcal L} }

\newcommand{\Lu}{\underline{L}}

\begin{document}

\begin{center}
{\bf \Large New Structures in Gravitational Radiation}
\end{center}

\begin{center}
{\bf Lydia Bieri \footnote{lbieri@umich.edu}} \\ \vspace{0.2cm}
{\itshape Dept. of Mathematics, University of Michigan, Ann Arbor, MI 48109-1120, USA}
\end{center}

\begin{abstract} 
We investigate the Einstein vacuum equations as well as the Einstein-null fluid equations describing neutrino radiation. 
We find new structures in gravitational waves and 
memory for asymptotically-flat spacetimes of slow decay. 
These structures do not arise in spacetimes resulting from data that is stationary outside a compact set. 
Rather the more general situations exhibit richer geometric-analytic interactions displaying the physics of these more general systems. 
It has been known that for stronger decay of the data, including data being stationary outside a compact set, gravitational wave memory is finite and of electric parity only. 
In this article, we investigate general spacetimes that are asymptotically flat in a rough sense. 
That is, the decay of the data to Minkowski space towards infinity is very slow. 
As a main new feature, we prove that there exists diverging magnetic memory sourced by the magnetic part of the curvature tensor (a) in the Einstein vacuum and (b) in the Einstein-null-fluid equations. The magnetic memory occurs naturally in the Einstein vacuum setting (a) of pure gravity. 
In case (b), 
in the ultimate class of solutions, the magnetic memory contains also a curl term from the energy-momentum tensor for neutrinos also diverging at the aforementioned rate. 
The electric memory diverges too, it is generated by the electric part of the curvature tensor and in the Einstein-null-fluid situation also by the corresponding energy-momentum component. 
In addition, we find a panorama of finer structures in these manifolds. Some of these manifest themselves as additional 
contributions to both electric and magnetic memory. 
Our theorems hold for any type of matter or energy coupled to the Einstein equations as long as the data decays slowly towards infinity and other conditions are satisfied. 
The new results have a multitude of applications ranging from mathematical general relativity to gravitational wave astrophysics, detecting dark matter and other topics in physics. 
\end{abstract}

\tableofcontents 

\section{Introduction}
\subsection{Overview and Summary of Results}

We find a wealth of new structures in gravitational waves and 
memory (a permanent change of the spacetime) for asymptotically-flat spacetimes of slow decay. In particular, we investigate 
the Einstein vacuum and Einstein-null-fluid equations describing neutrino radiation for sources whose distribution decays very slowly towards infinity. In particular, such sources are not stationary outside a compact set. For the physical case, we may assume the no-incoming radiation condition. We study the incoming radiation as well, then relate this to the situation where there is no incoming radiation. 
The new effects are expected to be seen in current and future gravitational wave detectors. 

First, we find growing magnetic memory within the realm of the Einstein vacuum (EV) equations. The electric memory is growing too. 
Then we show that all the new effects for the Einstein vacuum (EV) equations are also present in the Einstein-null fluid (ENF) situations. Whereas the magnetic memory is sourced by the corresponding magnetic component of the curvature tensor, the electric memory is sourced by the corresponding electric part of the curvature plus an integral term involving the shear. The null fluid contributes to the diverging electric memory at highest rate. 
In the ultimate class of solutions, the magnetic memory contains a curl term from the energy-momentum tensor for neutrinos also diverging at the aforementioned rate, and the integral of the shear term in the electric null memory becomes unbounded. 

The natural occurrence of magnetic memory in the EV and ENF setting for slowly decaying data is a unique feature of these more general spacetimes. If the decay of the data is stronger, which includes sources that are not stationary outside a compact set, then magnetic memory does not exist, and all the memory effects (being of electric parity only) are finite. 

Moreover, further interesting structures arise in our more general spacetimes. They will be derived and explored in detail in sections \ref{general},  \ref{inoutrad} and \ref{neutrinos} of this paper. 

Our new findings, in particular the new gravitational wave memories can be used to detect, identify and gain more information about sources. 

If dark matter behaves like neutrinos or similar matter that is not stationary outside a compact set, then the new phenomena 
can be used to detect dark matter via gravitational waves. 

\subsection{Spacetimes, Gravitational Radiation and Memory}
\label{slgrm}

The physical laws of the universe take a geometric form in the Einstein equations (\ref{ET}). They can be written as a system of second-order, hyperbolic, nonlinear (i.e. quasilinear) partial differential equations. The dynamics of the gravitational field are investigated via the Cauchy problem for physical initial data. The resulting spacetimes are solutions of the Einstein equations, respectively Einstein-matter equations if matter or energy fields are coupled to the original system. A long and interesting history of mathematical research has intertwined with the progress of General Relativity (GR). In particular, the mathematical endeavors of the pioneering years of GR culminated in Y. Choquet-Bruhat's proof of well-posedness of the Einstein equations \cite{ychb1} and her work with G. Geroch \cite{bruger}. These results have been generalized in many directions, and they form the foundations on which global results have been based. See Choquet-Bruhat's overview article for a more detailed discussion of this subject \cite{ychb3}. 

Gravitational waves are radiated away in extreme events such as binary black hole mergers binary neutron star mergers or core-collapse supernovae. Sources producing gravitational radiation are best described as asymptotically-flat spacetimes.

Gravitational waves in GR are predicted to change the spacetime permanently. This phenomenon is called the memory effect, and was found by Ya. Zel'dovich and B. Polnarev in a linearized theory \cite{zeldovich} and by D. Christodoulou in the fully nonlinear setting \cite{chrmemory}. There are two types of this memory \cite{lbdg3}. Namely, the present author and D. Garfinkle showed \cite{lbdg3} that the former by Zel'dovich and Polnarev, called ordinary memory, is 
related to fields that do not go out to null infinity, whereas the latter by Christodoulou, called null memory, is related to fields that do go out to null infinity.

For spacetimes decaying to Minkowski spacetime towards infinity at a rate of $r^{-1}$ or faster, all these memories  are finite and of electric parity only \cite{lydia4}. This includes sources that are stationary outside a compact set. 

In 2015, the first detection of gravitational waves from a binary black hole merger in the two LIGO facilities \cite{ligodetect1} marked a major scientific breakthrough. Another milestone followed in 2017 with LIGO and Virgo jointly measuring waves generated in a neutron star binary merger  \cite{ligodetect2, ligodetect3}. Many more events have been observed since 2015. We live in midst of the beginning of a new era where numerous gravitational wave detectors will reveal information from regions of our universe, where telescopes cannot see. 

P. Lasky, E. Thrane, Y. Levin, J. Blackman and Y. Chen suggest a way to detect gravitational 
wave memory with LIGO in \cite{Lasky1}. 

Whereas in detectors like LIGO gravitational memory will show as a permanent displacement of test masses, detectors like NANOGrav will recognize a frequency change of pulsars' pulses. 

The aforementioned pioneering works on the two types of gravitational wave memory \cite{zeldovich} and \cite{chrmemory} were followed by several contributions by other researchers \cite{blda1, blda2, braginsky, braginskyg, will, thorne, thorne2, jorg}. More recently, there has been an increasing number of authors contributing to the field \cite{1lpst1, 1lpst2, lbdg1, lbdg2, lbdg3,tolwal1,winicour, winma2, Lasky1, strominger,flanagan,favata,BGYmemcosmo1, bgsty1,twcosmo}. See the previous works for more detailed references. The literature on memory has grown very large, so that it is not feasible any longer to explain all  aspects in a research article. Rather we concentrate on the references that are relevant for the topics under investigation. Having said that, I recommend to explore the many aspects of memory in GR and their analogues in other theories. The aforementioned literature and references therein provide a comprehensive guidance. 

Garfinkle and the present author derived \cite{lbdg2} the analogues of both memories for the Maxwell equations. These were the first analogues outside GR. The search for analogues of the memory effect in other theories has become very popular. See for instance \cite{strominger}. 

Most physical matter- or energy-fields coupled to the Einstein equations contribute to the null memory \cite{lbdg3}. In particular, we showed this together with P.Chen and S.-T. Yau for the Einstein-Maxwell system, where a specific component of the electromagnetic field increases the null memory \cite{1lpst1}, \cite{1lpst2}. In collaboration with Garfinkle we proved that there is a contribution to the null memory from neutrino radiation \cite{lbdg1} as it occurs in a core-collapse supernova or a binary neutron star merger. 

In \cite{lydia4} the present author showed that for AF spacetimes approaching Minkowski spacetime at a rate of $r^{-1}$ and faster, memory is of electric parity only. An interesting and ``unusual" example of stress-energy of an expanding shell in linearized gravity was produced by G. Satishchandran and R. Wald \cite{WaldTm1} that gives rise to an ordinary magnetic memory. 

The present author also showed \cite{lydia4} that AF spacetimes approaching Minkowski spacetime at a rate of $o(r^{- \frac{1}{2}})$ generate diverging electric memory. 

It is interesting to point out the following: Whereas Satishchandran and Wald \cite{WaldTm1} invoked a very special example of stress-energy to produce magnetic memory, we find magnetic memory even in the Einstein vacuum regime (thus without any stress-energy present). We show that this new magnetic memory occurs naturally for slowly decaying AF spacetimes. 

In \cite{winma2}, based on Winicour's work \cite{winicour}, T. M\"adler and J. Winicour considered gravitational wave memory from different sources in a linearized setting. They showed that the special case of homogeneous, source-free gravitational waves coming in from past null infinity gives a magnetic memory, whereas all the other sources considered in their paper produce electric memory only. We can think of this as putting in magnetic memory by hand by placing these incoming waves at past null infinity. 
The results in \cite{winicour}, \cite{winma2} are consistent with the results from the afore-mentioned literature showing that all gravitational wave memory from systems decaying like $r^{-1}$ or faster are of electric parity only. In these settings, the only way that magnetic memory can occur is by having incoming radiation from past null infinity ``carry it" already. That is, it has to be in the initial data. 

In the present article, we show that magnetic memory arises naturally for systems that decay more slowly towards infinity. 
In particular, this includes the settings without incoming radiation. Thus, it is not contained in the initial data, but the evolution of the latter produce magnetic memory in the outgoing radiation. The situations with incoming radiation are discussed in detail in sections \ref{PNI} and \ref{beides} of this paper. 
In the latter, we give the full picture of incoming radiation, comprising gravitational waves of various decay properties, and explain the implications of turning on and off this incoming radiation. In this article, we prove that magnetic memory arises naturally in all these situations with and in particular without incoming radiation, thereby deriving new structures in gravitational waves and new memory.

Our results for the Einstein vacuum equations also hold for neutrino radiation via the Einstein-null-fluid equations. A further result in the present article is the derivation of a contribution to magnetic memory from neutrino radiation through a curl term of its stress-energy tensor in the ultimate class of spacetimes.

In GR, the dynamics of binary black holes, galaxies, generally of isolated gravitating systems (thus non-cosmological scenarios) are described by 
asymptotically-flat (AF) solutions of the Einstein vacuum equations, or Einstein equations coupled to corresponding matter or energy. These 
have been understood in detail (in the fully nonlinear regime) through the proofs of global nonlinear stability of Minkowski space. 
In the latter, we let small AF initial data (controlled via weighted Sobolev norms) evolve under the EV equations to become globally AF spacetimes that are causally geodescially complete (thus without any singularities). A semi-global result in the asymptotically-hyperbolic case was obtained by H. Friedrich \cite{fried1}. 
The first complete global proof was accomplished by D. Christodoulou and S. Klainerman in \cite{sta}. N. Zipser generalized this result to the Einstein-Maxwell system \cite{zip}, \cite{zip2}, and the present author in \cite{lydia1}, \cite{lydia2} to the borderline case for the EV equations assuming one less derivative and less fall-off by one power of $r$ than in \cite{sta} obtaining the borderline case in view of decay in power of $r$. Many authors contributed proofs or partial results in various directions. We would like to point out  \cite{vashin2017} for a more recent study of polyhomogeneous data. Please, see the aforementioned literature for more references on this topic. In this article, we concentrate on the works relevant to the problems under investigation. 
The results \cite{sta, zip, zip2, lydia1, lydia2}, while providing insights into large classes of important physical systems, they also establish a detailed description of null asymptotic behavior and gravitational radiation. The smallness of the initial data was required to establish existence and uniqueness of solutions. However, the main behavior along null hypersurfaces towards future null infinity remains largely independent from the smallness assumptions. 
To see this, one may first take a double-null foliation near scri. Next, consider the portion of null infinity 
near spacelike infinity such that the past of this portion intersected with the initial spacelike hypersurface lies within the large data region. 
Then find that the asymptotic results still hold for this portion of null infinity. 
This allows us to gain insights into large data behavior, including black holes. 

In this article, we first present the framework as well as a summary of our results, then proceed to the major setting and equations. In section \ref{inoutrad} we derive the new effects for the EV equations and in section \ref{neutrinos} for the Einstein-null fluid equations describing neutrino radiation. A comprehensive derivation of further new structures is provided. 
We conclude in part \ref{conclusions} with an outlook together with applications of the new phenomena.

\section{Main Structures and Main Results}
\label{settingresults}

The Einstein equations coupled to the stress-energy tensor of a general physical field are 
\be \label{ET}
G_{\mu \nu} \ := \ R_{\mu \nu} - \frac{1}{2} g_{\mu \nu} R \ = \ 8 \pi T_{\mu \nu}  \ , 
\ee 
for $\mu, \nu = 0,1,2,3$ and setting the constants $G=c=1$, where $G$ is Newton's gravitational constant, $c$ the speed of light. 
We solve the equations for the unknown metric $g_{\mu \nu}$. In (\ref{ET}), $G_{\mu \nu}$ denotes the Einstein tensor,  
$R_{\mu \nu}$ is the Ricci curvature tensor,  
$R$ the scalar curvature tensor, 
$T_{\mu \nu}$ is the energy-momentum tensor. 
The Einstein equations (\ref{ET}) have to be complemented with the corresponding equations for the field on the right hand side of (\ref{ET}). 

We denote the solution spacetimes by $(M,g)$. These are $4$-dimensional manifolds with a Lorentzian metric $g$ solving the system of equations. 

In this paper, Greek letters $\alpha, \beta, \gamma, \cdots$ denote spacetime indices, Latin letters $a, b, c, \cdots$ spatial indices, and capital Latin letters $A, B, C, \cdots$ indices on a two-dimensional spacelike surface diffeomorphic to $S^2$. When using a time foliation of the $4$-dimensional manifold into time $t$ and $3$-dimensional spacelike hypersurfaces, we may refer to the $0$-component as the $t$-component. 
For the larger part of our article we use a null foliation of the spacetime, that is introduced in 
section \ref{setting}.

The twice contracted Bianchi identities yield 
\be \label{AufdemBerg}
D^{\mu} G_{\mu \nu} = 0 \ . 
\ee
This implies that 
\be \label{AufdemFels}
D^{\mu} T_{\mu \nu} = 0 \ . 
\ee

Consider the Einstein vacuum equations 
\be \label{EV}
R_{\mu \nu} = 0  \ . 
\ee 

Consider the Einstein-null-fluid equations describing neutrino radiation in GR 
\be \label{ENF}
R_{\mu \nu}  \ = \ 8 \pi \ T_{\mu \nu} \ . 
\ee
As the $T_{\mu \nu}$ for the null fluid is traceless, the Einstein equations (\ref{ET}) for a null fluid reduce to (\ref{ENF}). 

The components of the energy-momentum tensor are given in section \ref{neutrinosintro}. 

Let $L$ denote an outgoing and $\underline{L}$ an incoming null vectorfield. They will be further specified in 
section \ref{setting}. 

Contract the contravariant tensor $T^{\mu \nu}$ with the metric to obtain the covariant tensor $T_{\mu \nu}$. Thus it is 
$T^{LL} = \frac{1}{4} T_{\underline{L} \underline{L}}$. \\ 

We investigate \\ 

$\bullet$ the {\bf Einstein vacuum equations (\ref{EV})} for large data consistent with \cite{lydia1, lydia2}, and \\ 

$\bullet$ the {\bf Einstein-null-fluid equations describing neutrino radiation in GR (\ref{ENF})} for data of the same general type together with a {\bf neutrino distribution} that {\bf decays slowly towards infinity}, in particular it is not stationary outside a compact set. \\ \\ 

{\bf Gravitational Wave Memory: Electric Parity, Magnetic Parity}. 

The Weyl tensor $W_{\alpha \beta \gamma \delta}$ is decomposed into its electric and magnetic parts, which are defined by
\begin{eqnarray}
{E_{ab}} := {W_{atbt}} \label{Wel1}
\\
{H_{ab}} := {\textstyle {1 \over 2}} {{\epsilon ^{ef}}_a}{W_{efbt}} \label{Wma1}
\end{eqnarray}
Here $\epsilon _{abc}$ is the spatial volume element and is related to the spacetime volume element by
${\epsilon_{abc}} = {\epsilon_{tabc}}$. The electric part of the Weyl tensor is the crucial ingredient in the 
equation governing the distance between two objects in free fall. In particular, their spatial separation $\Delta {x^a}$ is 
\begin{equation}
{\frac {{d^2}\Delta {x^a}} {d{t^2}}} = - {{E^a}_b}\Delta {x^b}
\label{geodev}
\end{equation}
A memory effect caused by the electric part of the curvature tensor is called {\bf electric parity memory} (i.e. electric memory); if it is caused by the magnetic part of the curvature tensor, then it is called {\bf magnetic parity memory} (i.e. magnetic memory). 
For asymptotically-flat (AF) systems with $O(r^{-1})$ decay towards infinity there is only {\bf electric parity memory}, {\bf magnetic parity memory} does not exist \cite{lydia4}. However, in AF spacetimes of slower decay, we show in this article that magnetic memory occurs  naturally. Moreover, the overall memory is growing and new structures arise that are not present in the former systems. These phenomena hold for the Einstein vacuum equations (\ref{EV}) as well as for the Einstein equations coupled to other fields (\ref{ET}) that {\bf decay slowly} towards infinity and obey other corresponding properties. In particular, they hold for the Einstein-null-fluid equations describing neutrino radiation (\ref{ENF}) for a neutrino distribution obeying the slow fall-off rates. \\

We use the notation introduced in \cite{sta} by D. Christodoulou and S. Klainerman. The spacetimes investigated in our article were treated by the present author in \cite{lydia1, lydia2}, where they obey {\bf smallness assumptions}. In this article, we study {\bf large data}. We emphasize that the main behavior along null hypersurfaces towards future null infinity remains largely independent from the smallness assumptions.

We are going to consider classes of asymptotically-flat initial data yielding corresponding classes of spacetimes. Let us first give the following definition. 

\begin{Def} (Christodoulou-Klainerman (CK), \cite{sta}) \label{intSAFCK} 
We define a strongly asymptotically flat initial data set in the sense 
of \cite{sta} and in the following denoted 
by (CK) initial data set, to be 
an initial data set $(H, \bar{g}, k)$, where 
$\bar{g}$ and $k$ are sufficiently smooth and there exists a coordinate system 
$(x^1, x^2, x^3)$ defined in a neighborhood of infinity such that, 
as 
$r = (\sum_{i=1}^3 (x^i)^2 )^{\frac{1}{2}} \to \infty$, 
$\bar{g}_{ij}$ and $k_{ij}$ are: 
\bea
\bar{g}_{ij} \ & = & \ (1 \ + \ \frac{2M}{r}) \ \delta_{ij} \ + \ o_4 \ (r^{- \frac{3}{2}}) \label{safg} \\ 
k_{ij} \ & = & \  o_3 \ (r^{- \frac{5}{2}}) \ ,  \label{safk}
\eea
where $M$ denotes the mass. 
\end{Def}
In \cite{sta}, the authors control weighted Sobolev norms of appropriate energies. 
This induces the above class of initial data.

\begin{Def} (Bieri (B), \cite{lydia1}, \cite{lydia2}) \label{intAFB} 
We define an asymptotically flat initial data set to be a 
(B) initial data set, if it is 
an asymptotically flat initial data set $(H_0, \bar{g}, k)$, 
where $\bar{g}$ and $k$ are sufficiently smooth 
and 
for which there exists a coordinate system $(x^1, x^2, x^3)$ in a neighborhood of infinity such that with 
$r = (\sum_{i=1}^{3} (x^i)^2 )^{\frac{1}{2}} \to \infty$, it is:  
\bea
\bar{g}_{ij} \ & = & \ \delta_{ij} \ + \ 
o_3 \ (r^{- \frac{1}{2}}) \label{afgeng}  \\
k_{ij} \ & = & \ o_2 \ (r^{- \frac{3}{2}})    \ .    \label{afgenk}  
\eea
\end{Def} 
In \cite{lydia1}, \cite{lydia2}, other weighted Sobolev norms of appropriate energies are controlled, 
yielding the most general class of spacetimes for which nonlinear stability has been proven. 

As a consequence from imposing less conditions on the data in \cite{lydia1}, \cite{lydia2}, the 
spacetime curvature is not in $L^{\infty}(M)$, as opposed to \cite{sta}. 
In \cite{lydia1}, \cite{lydia2}, only one derivative of the curvature (Ricci) lies in $L^2 (H)$. 

\begin{Def} (A) \label{intA} 
We define an asymptotically flat initial data set to be an (A) initial data set, if it is 
an asymptotically flat initial data set $(H_0, \bar{g}, k)$, 
where $\bar{g}$ and $k$ are sufficiently smooth 
and 
for which there exists a coordinate system $(x^1, x^2, x^3)$ in a neighborhood of infinity such that with 
$r = (\sum_{i=1}^{3} (x^i)^2 )^{\frac{1}{2}} \to \infty$, it is:  
\bea
\bar{g}_{ij} \ & = & \ \delta_{ij} \ + \ 
O_3 \ (r^{- \frac{1}{2}}) \label{Oafgeng}  \\
k_{ij} \ & = & \ O_2 \ (r^{- \frac{3}{2}})    \ .    \label{Oafgenk}  
\eea
\end{Def} 
Note the difference between definitions \ref{intAFB} and \ref{intA}. Namely, that little $o$ from the former is replaced by big $O$ in the latter. 

We introduce the following notation 

$\bullet$ {\bf (CKvac) Spacetimes} are solutions of the Einstein vacuum equations resulting from initial data as given in definition \ref{intSAFCK} with large data. 

$\bullet$ {\bf (Mvac) Spacetimes} are solutions of the Einstein vacuum equations resulting from initial data where in definition \ref{intSAFCK} the mass term $\frac{2M}{r}$ is replaced by a term being homogeneous of degree $-1$, the remainder of the metric $\bar{g}_{ij}$ decays like $r^{-1 - \epsilon}$ and $k_{ij}$ like $r^{-2 - \epsilon}$ towards infinity, with large data. 

$\bullet$ {\bf (Bvac) Spacetimes} are solutions of the Einstein vacuum equations resulting from initial data as given in definition \ref{intAFB} with large data. 

$\bullet$ {\bf (Avac) Spacetimes} are solutions of the Einstein vacuum equations resulting from initial data as given in definition \ref{intA} with large data. 

$\bullet$ {\bf (B-Tneutrinos) Spacetimes} are solutions of the Einstein-null-fluid equations describing neutrino radiation for initial data as given in definition \ref{intAFB} and the corresponding energy-momentum tensor, which is {\bf falling off at slow rates} (specified in section \ref{neutrinos}), large data. In particular, such sources are not stationary outside a compact set. 

$\bullet$ {\bf (A-Tneutrinos) Spacetimes} are solutions of the Einstein-null-fluid equations describing neutrino radiation for initial data as given in definition \ref{intA} and the corresponding energy-momentum tensor, which is {\bf falling off at slow rates} (specified in section \ref{neutrinos}), large data. In particular, such sources are not stationary outside a compact set. \\ 

{\bf Stability Theorems:} It is important to emphasize that for data as in definition \ref{intAFB} under a smallness condition, the present author established \cite{lydia1}, \cite{lydia2} a global existence and decay theorem for the Einstein vacuum equations  (\ref{EV}). When the smallness condition is relaxed and we allow for large data, then recall from the end of section \ref{slgrm}, that there exists a complete domain of dependence of the complement of a sufficiently large compact subset of the initial hypersurface. That is, we have a solution spacetime with a portion of future null infinity corresponding to all values of the retarded time $u$ not greater than a fixed constant. This provides the solid foundation to investigate the asymptotic behavior at future null infinity for large data, that is for (Bvac) spacetimes. In particular, it allows us to prove theorems on the nature of gravitational radiation for (Bvac) spacetimes. Naturally, our investigations will extend to (B-Tneutrinos) spacetimes. Note that for data of type (B) the total energy is finite but the total angular momentum diverges. If we make the class of initial data even larger to be represented by the type (A) as in definition \ref{intA}, then for this class (A) the total energy is no longer finite. However, for type (A) data no existence theorem is known for a development which includes a portion of future null infinity. Therefore, the study of (Avac) and (A-Tneutrinos) spacetimes will lead to conjectures furnished with supporting evidence. In contrast, for data yielding (CKvac) and (Mvac) spacetimes, not only the total energy but also the total angular momentum is finite.  \\

In the following, we will often use the abbreviations: IR for ``incoming radiation", NIR for ``no incoming radiation", 
and NIRC for ``no incoming radiation condition". \\ 

We state our results for the Einstein vacuum equations (\ref{EV}) and for the Einstein-null-fluid equations (\ref{ENF}) describing neutrino radiation in GR. 
Investigating the properties of gravitational radiation for very general spacetimes, we show: \\ 

For the {\bf Einstein vacuum equations (\ref{EV})}: 
\begin{itemize}
\item[\bf{(1)}] {\bf (Bvac) spacetimes}: Initial data as given in definition \ref{intAFB} for large data lead to the following types of {\bf memory effects at future null infinity $\mathcal{I}^+$}: 

{\bf (a)} the {\bf electric memory} effect {\bf growing at rate $\sqrt{|u|}$} derived by the present author in \cite{lydia4}, which is sourced by the corresponding {\itshape electric part of the curvature} growing at the same rate and a {\itshape finite contribution from the shear (the news tensor)} of the outgoing radiation; 

{\bf (b)} a {\bf new magnetic memory} effect {\bf growing at rate $\sqrt{|u|}$}, sourced by the corresponding {\itshape magnetic part of the curvature} growing at the same rate.

\item[\bf{(2)}] {\bf Initial data as in definition \ref{intAFB}} for large data lead to the following types of {\bf memory effects at past null infinity $\mathcal{I}^-$}: 

{\bf (a)} an {\bf electric memory growing at rate $\sqrt{ | \underline{u} | }$}, which is sourced by the corresponding {\itshape electric part of the curvature} growing at the same rate and a {\itshape finite contribution from the shear} of the incoming radiation; 
 
{\bf (b)} a {\bf magnetic memory growing at rate $\sqrt{ | \underline{u} | }$}, which is sourced by the corresponding {\itshape magnetic part of the curvature} growing at the same rate. 

\item[{\bf (3)}] If we impose the {\bf no incoming radiation condition} at past null infinity $\mathcal{I}^-$, then both {\bf effects (2)(a) and (2)(b) at $\mathcal{I}^-$  disappear}, that is the corresponding limits in the relevant equations are zero. 
These spacetimes, obeying the no incoming radiation condition, form a subclass of the spacetimes generated by the past evolution of large initial data as in \cite{lydia1, lydia2} (without any smallness assumptions). 
We find that data with and without the NIRC at $\mathcal{I}^-$ obey the same energy conditions. 
\item[{\bf (4)}] Conjecture: For {\bf (Avac) spacetimes}, all the above results hold with the difference that the null memory may not be bounded anymore in (Avac). 
 
\end{itemize}

For the {\bf Einstein-null-fluid equations (\ref{ENF}) describing neutrino radiation in GR}: 
\begin{itemize}
\item[\bf{(5)}] {\bf (B-Tneutrinos) Spacetimes} feature the following types of {\bf memory effects at future null infinity $\mathcal{I}^+$}: 

{\bf (a)} an {\bf electric memory} effect {\bf growing at rate $\sqrt{|u|}$} sourced by the corresponding {\itshape electric part of the curvature} 
and {\itshape the $T_{\underline{L} \underline{L}}$ component of the energy-momentum tensor}, 
each growing at the same rate, and a {\itshape finite contribution from the shear} ; 

{\bf (b)} a {\bf new magnetic memory} effect {\bf growing at rate $\sqrt{|u|}$}, sourced by the corresponding {\itshape magnetic part of the curvature} growing at the same rate. 

\item[\bf{(6)}] Conjecture: {\bf (A-Tneutrinos) Spacetimes} produce memories like in (5(a)-(b)) but none of them are bounded anymore. In addition, a {\bf new} integral of a curl of $T$ term, growing like $\sqrt{|u|}$ contributes to to the magnetic memory. 

\end{itemize}

For {\bf both} systems: {\bf EV (\ref{EV}) as well as ENF equations (\ref{ENF}) }: 
\begin{itemize}
\item[\bf{(7)}] In all the spacetimes {\bf (Bvac), (B-Tneutrinos)} finer {\bf structures} in the curvature as well as shear terms appear in both electric and magnetic memory. Whereas the ones related to curvature contribute at diverging and finite levels, the new structures in the shear expressions yield additional finite memories. 

\item[\bf{(8)}] Conjecture: In the spacetimes {\bf (Avac), (A-Tneutrinos)}, similar as described in the previous point, finer structures arise. 
\end{itemize}

The results for the (B-Tneutrinos) spacetimes and the conjectures for (A-Tneutrinos) spacetimes {\bf hold for any Einstein-matter system with an energy-momentum tensor behaving correspondingly}. Necessary yet not sufficient: the latter obeys slow decay laws. In particular, such sources are not stationary outside a compact set.   \\ 

The main results of this article are given in theorems \ref{theelmagn***} as well as \ref{theneutrinoscurl55}, and the investigations of this paper lead us to state conjectures \ref{conjtheelmagn***} and \ref{conjneutrinoscurl55}. \\

{\bf Sources: non-stationary versus stationary outside a compact set:} In \cite{lbdg1} Garfinkle and the present author investigated neutrino radiation in GR via the Einstein-null fluid equations for sources in spacetimes falling off like (\ref{safg}). As a consequence the stress-energy of these sources enjoys a fast decay behavior. In particular, this includes sources that are stationary outside a compact set. 
We found that all the memories are finite, and that the neutrinos contribute to the null memory by a finite amount. No magnetic memory is present. The new situation investigated in the current article is not stationary outside a compact set but enjoys very slow decay towards infinity. This very fact causes not only diverging memories but also new structures to play more dominant roles. The emergence of magnetic memory is a most striking feature of the more general sources. It is missing completely for sources of stronger decay. In this article, we derive magnetic memory for Einstein-null-fluid systems describing neutrino distributions of slow decay as in 
(B-Tneutrinos) spacetimes, and we conjecture corresponding structures for (A-Tneutrinos) spacetimes. As a side result, we show as well that spacetimes solving equations (\ref{ENF}) 
 and that are falling off at the order $O(r^{-1})$ without any assumptions on the leading order decay do not produce any magnetic memory.

\section{Setting}
\label{setting}

In this section, we set up the stage to derive the new effects.

The covariant differentiation on the spacetime $M$ is written as $D$ or $\nabla$, and the one on a spacelike hypersurface $H$ is $\overline{\nabla}$ or $\nabla$. It is clear from the context, what $\nabla$ refers to.

Let $t$ be the maximal time function and $u$ the optical function as in \cite{lydia1, lydia2}. The former foliates the spacetime into spacelike hypersurfaces $H_{t}$, the latter into outgoing null hypersurfaces $C_u$. 
Moreover, we denote the intersection between these hypersurfaces by $S_{t,u} := H_t \cap C_u$. 
The surfaces $S_{t,u}$ are diffeomorphic to the sphere $S^2$ and we will refer to $\theta_1, \theta_2$ given on $S_{t,u}$ as the spherical variables. 
We also introduce $\underline{u} := - u + 2r$ with $r = r(t,u)$ being defined by $4 \pi r^2$ expressing the surface area of $S_{t,u}$. 

Naturally, we denote by $\underline{C}_{\underline{u}}$ the incoming null hypersurfaces. 
Define $\tau_- := \sqrt{1 + u^2}$, and $\tau_+ := \sqrt{1 + \underline{u}^2}$. 
We work with a null frame $e_1, e_2, e_3, e_4$, where $\{ e_A \}, A = 1,2$ is a local frame field for $S_{t,u}$, and $e_3 = \underline{L}, e_4 = L$ are a null pair. 
In particular we have $g(e_4, e_3) = -2$. 
We define the tensor of projection from the tangent space of $M$ to that of $S_{t,u}$ by 
\[
\Pi^{\mu \nu} = g^{\mu \nu} + \frac{1}{2} ( e_4^{\nu} e_3^{\mu} + e_3^{\nu} e_4^{\mu} )
\]
Further, let $T$ be the future-directed unit normal to $H_t$, and $N$ the outward unit normal to $S_{t,u}$ in $H_t$. Then we see that the outgoing null vector field $e_4 = T+N$, and the incoming null vector field $e_3 = T-N$. 
Operators on the surfaces $S_{t,u}$ will be denoted with a slash. Thus, $\nlap$ describes covariant differentiation on $S_{t,u}$, the divergence and curl operators for tensors $t$ on $S_{t,u}$ are given as 
$\dlap t_{A_1 \cdots A_k} = \nlap^B t_{A_1 \cdots A_k B}$ and 
$\clap \ \  t_{A_1 \cdots A_k} = \epsilon^{BC} \nlap_B t_{A_1 \cdots A_k C}$, respectively. 
Moreover, for a $p$-covariant tensor field $t$ tangent to $S$, $D_4t$ and $D_3t$ are the projections to $S$ of the Lie derivatives 
$\Lie_4 t$, respectively $\Lie_3 t$.

We decompose the second fundamental form $k_{ij}$ of $H_t$ into 
\bea
k_{NN} &=& \delta \\
k_{AN} &=& \epsilon_A\\
k_{AB} &=& \eta_{AB}  \ \ . 
\eea
Define
\be \label{theta1}
  \theta_{AB} =  \langle \nabla_{A} N, e_B \rangle.
\ee

\begin{Def}
We define the null components of the Weyl curvature $W$ as follows: 
\bea
\underline{\alpha}_{\mu \nu} \ (W) \ & = & \ 
\Pi_{\mu}^{\ \rho} \ \Pi_{\nu}^{\ \sigma} \ W_{\rho \gamma \sigma \delta} \
e_3^{\gamma} \ e_3^{\delta} 
\label{underlinealpha} \\ 
\underline{\beta}_{\mu} \ (W) \ & = & \ 
\frac{1}{2} \ \Pi_{\mu}^{\ \rho} \ W_{\rho \sigma \gamma \delta} \  e_3^{\sigma} \
e_3^{\gamma} \ e_4^{\delta} 
\label{underlinebeta} \\ 
\rho \ (W) \ & = & \ 
\frac{1}{4} \ W_{\alpha \beta \gamma \delta} \ e_3^{\alpha} \ e_4^{\beta} \
e_3^{\gamma} \ e_4^{\delta} 
\label{rho} \\ 
\sigma \ (W) \ & = & \ 
\frac{1}{4} \ \ ^*W_{\alpha \beta \gamma \delta} \ e_3^{\alpha} \ e_4^{\beta} \
e_3^{\gamma} \ e_4^{\delta} 
\label{sigma} \\ 
\beta_{\mu}  \ (W) \ & = & \  
\frac{1}{2} \ \Pi_{\mu}^{\ \rho} \ W_{\rho \sigma \gamma \delta} \ e_4^{\sigma} \
e_3^{\gamma} \ e_4^{\delta} 
\label{beta} \\ 
\alpha_{\mu \nu} \ (W) \ & = & \ 
\Pi_{\mu}^{\ \rho} \ \Pi_{\nu}^{\ \sigma} \ W_{\rho \gamma \sigma \delta} \
e_4^{\gamma} \ e_4^{\delta}  \ . 
\label{alphaR}
\eea
\end{Def}

Thus the following holds, where capital indices take the values $1,2$: 
\bea
W_{A3B3} \ & = & \ \underline{\alpha}_{AB} \label{intnullcurvalphaunderline*1} \\ 
W_{A334} \ & = & \ 2 \ \underline{\beta}_A \\ 
W_{3434} \ & = & \ 4 \ \rho \\ 
\ ^* W_{3434} \ & = & \ 4 \ \sigma \\ 
W_{A434} \ & = & \ 2 \ \beta_A \\ 
W_{A4B4} \ & = & \ \alpha_{AB}  \label{intnullcurvalpha*1}
\eea
with \\ 
\begin{tabular}{lll}
$\alpha$, $\underline{\alpha}$ & : & $S$-tangent, symmetric, traceless tensors \\ 
$\beta$, $\underline{\beta}$ & : &  $S$-tangent $1$-forms \\ 
$\rho$, $\sigma$ & : & scalars \ . \\ 
\end{tabular} 
\\ \\ 
Whenever we work with the Einstein vacuum equations, the Riemannian curvature tensor $R_{\alpha \beta \gamma \delta}$ is identically the Weyl curvature tensor $W_{\alpha \beta \gamma \delta}$. 

Let us introduce the shears $\widehat{\chi}$, $\underline{\widehat{\chi}}$ to be the traceless parts of the second fundamental forms 
with respect to the null vectorfields $L$ and $\underline{L}$ generating the corresponding outgoing, respectively incoming null hypersurfaces (``light cones"). 
Let $X, Y$ be arbitrary tangent vectors to $S_{t, u}$ at a point in this surface. Then
the second fundamental forms are defined to be 
\[
\chi (X, Y) = g(\nabla_X L, Y) \ \, \ \ \underline{\chi} (X, Y) = g(\nabla_X \underline{L}, Y) . 
\]
We denote the trace of these tensors by $tr \chi$, respectively $tr \underline{\chi}$. 
Complementing the above, the Ricci rotation coefficients of the null frame are: 
\beas
\chi_{AB} & = & g(D_A e_4, e_B)  \\ 
\underline{\chi}_{AB} & = & g(D_A e_3, e_B)  \\ 
\underline{\xi}_A & = & \frac{1}{2} g(D_3 e_3, e_A)  \\ 
\zeta_A & = & \frac{1}{2} g(D_3 e_4, e_A)  \\ 
\underline{\zeta}_A & = & \frac{1}{2} g(D_4 e_3, e_A)  \\ 
\nu & = & \frac{1}{2} g(D_4 e_4, e_3)  \\ 
\underline{\nu} & = & \frac{1}{2} g(D_3 e_3, e_4)  \\ 
\epsilon_A & = & \frac{1}{2} g(D_A e_4, e_3)
\eeas
Here, $\zeta$ is the torsion-one-form.

Next, we introduce a useful concept and notation that we will use extensively later in this paper. 
The {\itshape signature} $s$ is defined to be the difference of the number of contractions 
with $e_4$ minus the number of contractions with $e_3$. 
Then we introduce the following.

\begin{Def}
Let $W$ be an arbitrary Weyl tensor and let $\xi$ be any of its null components. 
Let $\Dlap_3 \xi$ and $\Dlap_4 \xi$ denote the projections to $S_{t,u}$ of $D_3 \xi$ and $D_4 \xi$, respectively. 
Define the following $S_{t,u}$-tangent tensors: 
\beas
\xi_3 \ & = & \ \Dlap_3 \xi \ + \ \frac{3-s}{2} tr \underline{\chi} \xi \\ 
\xi_4 \ & = & \ \Dlap_4 \xi \ + \ \frac{3+s}{2} tr \chi \xi  \ \ . 
\eeas 
\end{Def}

For the Einstein vacuum equations we find that 
the energy radiated away per unit angle in a given direction is $F/4\pi$ with 
\be \label{F1}
F(\cdot) = \frac{1}{2} \int_{- \infty}^{+ \infty}  \mid \Xi(u, \cdot) \mid^2  du  \ 
\ee 
and $\Xi$ denoting the corresponding limit of the shear. 

For the Einstein-null fluid equations describing spacetimes with neutrinos 
we find that the energy radiated away per unit angle in a given direction is $F_T/4\pi$ with 
\be \label{FXiT1}
F_T(\cdot) = \frac{1}{2} \int_{- \infty}^{+ \infty} \left( \mid \Xi(u, \cdot) \mid^2 \ + \ 2 \pi \ \mathcal{T}_{33} (u, \cdot) \right) du  \ . 
\ee 
In the special class of (A-Tneutrinos) spacetimes we find that the angular momentum radiated away caused by the matter is 
\be \label{RcurlT1}
\mathcal{A}_T (\cdot) = 4 \pi \int_{- \infty}^{+ \infty}  \big{(} \clap \ \ T \big{)}^*_{34_3} (u, \cdot)  du  \ . 
\ee
Hereby, $\mathcal{T}_{33}$ and $\big{(} \clap \ \ T \big{)}^*_{34_3}$ denote corresponding limits generated by the stress-energy tensor of matter, that is the null fluid describing the neutrino distribution. 

The details will be derived in the following sections.

\section{General Spacetimes}
\label{general}

\subsection{Overview}

Let us have a closer look at (Bvac) spacetimes. 
That is, we have data as in \cite{lydia1}, \cite{lydia2} but the data is not required to be small. Thus, we allow for large data. First, we recall structures from \cite{lydia1}, \cite{lydia2}. 
There, it is shown that in the global future development of the initial data, the following terms have a decay behavior at infinity that is given by 
\bea
\underline{\alpha} \ & = & \ O \ ( r^{- 1} \ \tau_-^{- \frac{3}{2}})   \label{resua1} \\ 
\underline{\beta} \ & = & \ O \ ( r^{- 2} \ \tau_-^{- \frac{1}{2}})  \label{resub1}   \\ 
\rho , \ \sigma , \ \alpha , \ \beta \ & = & \ o \ (r^{- \frac{5}{2}})   \label{resrsab1}   \\ 
\hat{\chi}  \ & = & \ o \ (r^{- \frac{3}{2}})   \label{reshatchi1}   \\ 
\underline{\hat{\chi}}  \ & = & \ O \ (r^{-1} \tau_-^{- \frac{1}{2}})  \label{resuhatchi1}  \\ 
tr \chi - \overline{tr \chi} \ & = & \ O \ (r^{- 2})   \label{restrchi1}  \\ 
\zeta  \ & = & \ o \ (r^{- \frac{3}{2}})  \label{resz1}   \\ 
\underline{\zeta}  \ & = & \ o \ (r^{- \frac{3}{2}})  \label{resuz1}   \\ 
K  - \frac{1}{r^2}  \ & = & \  o \ (r^{- \frac{5}{2}})   \label{resK1} \\ 
tr \chi \ & = & \  \frac{2}{r}  \ + \ l.o.t. \label{trchi**1} \\ 
tr \underline{\chi} \ & = & \  - \frac{2}{r} \ + \ l.o.t.  \label{trchibar**1} 
\eea
with $K$ the Gauss curvature of the surfaces $S_{t,u}$. 
Further, we have 
\beas
\theta \ & = & \ 
O(r^{-1} \tau_-^{- \frac{1}{2}}) \\ 
\hat{\eta}  \ & = & \ 
O(r^{-1} \tau_-^{- \frac{1}{2}}) \\ 
\epsilon \ & = & \ o(r^{- \frac{3}{2}})  \\ 
\delta \ & = & \ o(r^{- \frac{3}{2}})  \\ 
\eeas

The mass aspect function $\mu$ and its conjugate $\underline{\mu}$ are defined as 
\bea
\mu & = & K + \frac{1}{4} tr \chi tr \underline{\chi} - \dlap \zeta  \label{mu} \\
\underline{\mu} & = & K + \frac{1}{4} tr \chi tr \underline{\chi} + \dlap \zeta  \label{umu} 
\eea

Moreover, it is proven in \cite{lydia1}, \cite{lydia2} that the mass aspect function $\mu = O(r^{-3})$. We use propagation equations to prove this result, thereby the leading order terms cancel. 

\subsection{Structures}

We consider the propagation equations 
\bea
\frac{d}{ds} \hat{\chi} + tr \chi \hat{\chi} \ & = & \ \alpha  \label{schihat1}  \\ 
\Dlap_3 \hat{\chi} + \frac{1}{2} tr \underline{\chi} \hat{\chi} \ & = & \ 
- 2 \omega \hat{\chi} -  \frac{1}{2} tr \chi \underline{\hat{\chi}} +  \nlap \hat{\otimes} \zeta + \zeta \hat{\otimes} \zeta  \label{D3chihat1}  \\ 
\frac{d}{ds} \underline{\hat{\chi}} + tr \chi { \underline{\hat{\chi}}} \ & = & \ 
- \frac{1}{2} tr \underline{\chi} \hat{\chi} + \nlap \hat{\otimes} \underline{\zeta} +  \underline{\zeta} \hat{\otimes}  \underline{\zeta}   \label{suchihat1}  \\ 
\Dlap_3  \underline{\hat{\chi}} \ & = & \ 
- \underline{\alpha}  \label{D3uchihat1}  \\ 
\frac{d}{ds}  tr \underline{\chi} + \frac{1}{2} tr  \chi tr \underline{\chi} \ & = & \ 
2 \rho + 
2 \dlap \underline{\zeta} 
- \hat{\chi} \cdot \underline{\hat{\chi}} + 2 | \underline{\zeta} |^2  \label{struchi1} \\ 
\ & = & \ 
- 2 \underline{\mu} + 2 | \underline{\zeta} |^2 \nonumber  \\ 
\Dlap_3 \zeta + tr \underline{\chi} \zeta \ & = & \ 
- \underline{\beta} - 2 \underline{\hat{\chi}} \cdot \zeta + 2 \nlap \omega + \frac{1}{2} tr \chi \underline{\xi}  + \hat{\chi} \cdot \underline{\xi}  \label{D3zeta1}  
\eea
\bea
D_3 tr \chi \ + \ \frac{1}{2} tr \underline{\chi} tr \chi \ & = & \ 
- 2 \omega tr \chi \ + \ 2 \dlap \zeta - \hat{\chi} \cdot \hat{\underline{\chi}} \ + \ 2 |\zeta|^2 \ + \ 2 \rho  \label{D3trchi1} \\ 
\ & = & \ 
- 2 \omega tr \chi - 2 \mu + 2 | \zeta |^2 \nonumber \\ 
\frac{d}{ds} tr \chi \ + \ \frac{1}{2} (tr \chi)^2 \ & = & \ 
- | \hat{\chi} |^2  \label{strchi1} 
\eea
where $\omega$ is another Ricci coefficient at the order of $\delta$.

Let $K$ be the Gauss curvature of $S_{t,u}$. 
The Gauss equation reads 
\be \label{Gauss1}
 K + \frac{1}{4} tr \chi tr \underline{\chi} - \frac{1}{2} \hat{\chi} \cdot \hat{\underline{\chi}} = - \rho 
\ee

The shears $\hat{\chi}$ and $\underline{\hat{\chi}}$ obey the equations 
\bea
\dlap \hat{\underline{\chi}} & = & \underline{\beta} + \hat{\underline{\chi}} \cdot \zeta 
+ \frac{1}{2} ( \nlap tr \underline{\chi} - tr \underline{\chi} \zeta ) \ = \   \underline{\beta} + l.o.t. \label{dchibund1} \\ 
\dlap \hat{\chi} & = & - \beta - \hat{\chi} \cdot \zeta + \frac{1}{2} ( \nlap tr \chi + tr \chi \zeta )  \label{dchib1}
\eea
Recall that $\zeta$ is the torsion-one-form. 

The Hodge system for $\zeta$ on $S_{t,u}$ is 
\bea
\dlap \zeta & = & - \mu - \rho +  \frac{1}{2} \hat{\chi} \cdot \hat{\underline{\chi}}  \label{dzeta} \\ 
\clap \ \  \zeta & = & \sigma -  \frac{1}{2} \hat{\chi} \wedge \hat{\underline{\chi}}  \label{czeta}
\eea
where $\mu$ is given in (\ref{mu}), $\underline{\mu}$ in (\ref{umu}).  

The shears are related to each other by the equation 
\be \label{chihat}
\frac{\partial}{\partial u} \hat{\chi} \ = \ \frac{1}{4} tr \chi \cdot \hat{\underline{\chi}} + l.o.t. 
\ee

Also, it is 
\be \label{uchihat}
\frac{\partial}{\partial u} \underline{\hat{\chi}} \ = \ \frac{1}{2} \underline{\alpha} + l.o.t. 
\ee

Note that (\ref{D3chihat1}) yields (\ref{chihat}), and (\ref{D3uchihat1}) yields (\ref{uchihat}).

\subsection{Limits at Null Infinity}
\label{limits12} 
 
{\bf Limits at null infinity $\mathcal{I^+}$:}  
Next, we are going to describe a very interesting phenomenon revealing the dynamical and non-dynamical parts of the geometric components. 

As a result of the proof in \cite{lydia1, lydia2}, several quantities, which are defined locally on the surface $S_{t,u}$, do not attain corresponding limits on a given null hypersurface $C_u$ as 
$t \to \infty$. However, the difference of their values at corresponding points on $S_{u}$ and $S_{u_0}$ does tend to a limit. In particular, let us look at $\hat{\chi}$ that is defined locally on $S_{t,u}$. Recall (\ref{reshatchi1}). 
Even though $r^2 \hat{\chi}$ does not have a limit as $r \to \infty$ on a given $C_u$, the difference at corresponding points on $S_u$ in 
$C_u$ and on $S_{u_0}$ in $C_{u_0}$ does have a limit. 
In particular, these points being joined by an integral curve of $e_3$, the said difference attains the limit 
\be \label{limitsconcept1}
\int_{u_0}^{u} \Dlap_3 \hat{\chi} \ du' 
\ee
The part of $\hat{\chi}$ with slow decay of order $o(r^{- \frac{3}{2}})$ is non-dynamical, that is, it does not evolve with $u$. We see that this part does not tend to any limit at null infinity $\mathcal{I^+}$. 
Similarly, the components of the curvature that are not peeling have leading order terms that are non-dynamical (and do not attain corresponding limits at $\mathcal{I^+}$). Taking off these pieces gives us the dynamical parts of these (non-peeling) curvature components.

{\itshape Notation:} In (Bvac) and (B-Tneutrinos) spacetimes, we denote the part of $\hat{\chi}$ with decay $o(r^{- \frac{3}{2}})$ and which is non-dynamical (i.e. which does not evolve with $u$) by $[r^{- \frac{3}{2}}]$. Denote the leading order dynamical part of $\hat{\chi}$ (i.e. which evolves with $u$) by $\{ r^{-2}  \tau_-^{+ \frac{1}{2}} \}$. More generally, for any of the non-peeling curvature components and any of the Ricci coefficients which have a leading order non-dynamical part, let $[ \cdot ]$ denote the leading order non-dynamical part (thus not evolving in $u$) of this component; and let $\{ \cdot \}$ denote its leading order dynamical part (thus evolving in $u$).

The corresponding notation will be used for (Avac) and (A-Tneutrinos) spacetimes, with the difference that the non-dynamical part of  $\hat{\chi}$ is of order $O(r^{- \frac{3}{2}})$, and more generally for the other quantities the non-dynamical parts will be of corresponding order $O(\cdot)$ rather than $o(\cdot)$.

{\itshape Dynamical versus non-dynamical:} Given the previous argument together with the proof in \cite{lydia1}, \cite{lydia2}, the following is a consequence of equations (\ref{chihat})-(\ref{uchihat}) for both {\bf small as well as large data}: 
\bea
\hat{\chi} & = & [r^{- \frac{3}{2}}] + \{ r^{-2} \tau_-^{+ \frac{1}{2}} \}  + l.o.t. \label{hchiterms1} \\ 
\underline{\hat{\chi}} & = &  \{ r^{-1} \tau_-^{- \frac{1}{2}} \} + [r^{- \frac{3}{2}}]+ l.o.t.  \label{huchiterms1}
\eea

Further, by the proof in \cite{lydia1}, \cite{lydia2} and the {\bf smallness conditions} therein for the $e_3$-derivative of $\rho$, respectively $\sigma$: 
\be \label{rho3bound}
\int_H r^4 |\rho_3|^2 \leq c \epsilon 
\ee
\be \label{sigma3bound}
\int_H r^4 |\sigma_3|^2 \leq c \epsilon 
\ee
it is a consequence that 
\[
\rho_3 = O(r^{-3} \tau_-^{- \frac{1}{2}}) \ \ \ , \ \ \ \sigma_3 = O(r^{-3} \tau_-^{- \frac{1}{2}})
\]
Thus, $\rho_3$, respectively, $\sigma_3$ cannot have any terms of the order $r^{- \frac{5}{2}} \tau_-^{- \frac{3}{2}}$. 
In particular, the curvature components $\rho$ and $\sigma$ have the following structures, where we use the notation from above: 
\bea
\rho & = & [ r^{- \frac{5}{2}} ] + \{ r^{-3}  \tau_-^{+ \frac{1}{2}}  \} + \{ r^{- 3} \}  \label{rhoterms1} 
+ \{ r^{-3}  \tau_-^{+ \beta}  \} 
+  O(r^{-3} \omega^{-\alpha})  
\eea
and 
\bea
\sigma & = & [ r^{- \frac{5}{2}} ] + \{ r^{-3}  \tau_-^{+ \frac{1}{2}}  \} + \{ r^{- 3} \} \label{sigmaterms1} 
+ \{ r^{-3}  \tau_-^{+ \beta}  \} 
+  O(r^{-3} \omega^{-\alpha}) 
\eea
with $\omega$ denoting $r$ or $\tau_-$ and $\alpha > 0$, $0 < \beta < \frac{1}{2}$. 
We emphasize that (\ref{rhoterms1})-(\ref{sigmaterms1}) are consequences of the smallness assumption on the initial data and the proof in  \cite{lydia1}, \cite{lydia2}. For {\bf large} data, there are more terms present which feature a large variety of decay, possibly including terms in $\rho$, respectively $\sigma$, of the order 
$r^{- \frac{5}{2}} \tau_-^{- \alpha}$ with $\alpha > 0$. Nevertheless, the highest order terms will behave in the same ways.

A direct consequence of the results of \cite{lydia1, lydia2} is the following 
\begin{The} \label{nummer1}
For the spacetimes of types (Bvac) and (B-Tneutrinos), the normalized curvature components $r\underline{\alpha }\left( W\right) $, $r^{2}\underline{\beta }\left(
W\right) $ have limits on $C_u$ as $t\rightarrow \infty $: 
\begin{eqnarray*}
\lim_{C_{u},t\rightarrow \infty }r\underline{\alpha }\left( W\right)
&=&A_{W}\left( u,\cdot \right) ,\, \ \ \ \ \ \ \ \ \ \ \ \
\lim_{C_{u},t\rightarrow \infty }\,r^{2}\underline{\beta }\left( W\right)
= \underline{B}_{W}\left( u,\cdot \right) \ , 
\end{eqnarray*}
where the limits are on $S^{2}$ and depend on $u$. These limits satisfy 
\begin{eqnarray*}
\left| A_{W}\left( u,\cdot \right) \right| &\leq &C\left( 1+\left| u\right|
\right) ^{-3/2}\, \, \ \ \ \ \ \ \ \ \ \ \ \ \left| \underline{B}_{W}\left( u,\cdot
\right) \right| \leq C\left( 1+\left| u\right| \right) ^{-1/2}  . \ \
\end{eqnarray*}
\end{The}
Moreover, the following limit exists 
\be \label{limXiThe1}
- \frac{1}{2} \lim_{C_{u},t\rightarrow \infty }r\widehat{\underline{\chi }} = \lim_{C_{u},t\rightarrow \infty } r \hat{\eta} = 
\Xi \left( u,\cdot \right) 
\ee 
Further, it follows from (\ref{uchihat}), respectively from (\ref{dchibund1}) that 
\begin{eqnarray}
\frac{\partial \Xi }{\partial u} &=&-\frac{1}{4}A_{W}  \label{1XiAT1} \\ 
\underline{B} &=& - 2 \dlap \Xi \label{Ldchibund1}
\end{eqnarray}

\begin{conj} \label{Conjnummer1}
For the spacetimes of types (Avac) and (A-Tneutrinos), the corresponding statements of theorem \ref{nummer1} hold. Moreover, (\ref{limXiThe1})-(\ref{Ldchibund1}) hold. 
\end{conj}

\section{Incoming and Outgoing Radiation}
\label{inoutrad}

In section \ref{inoutrad}, we investigate spacetimes of type (Bvac) as well as (Avac). 

The outgoing radiation is dominated by $\underline{\hat{\chi}}$ and the incoming radiation by $\hat{\chi}$. 

The energy in form of incoming gravitational waves is defined at past null infinity. 
In \cite{DCblh2008}, Demetrios Christodoulou replaced this notion by an integral over advanced time of  
\be \label{en}
e := \frac{1}{2} | \hat{\chi} |^2 
\ee
In \cite{DCblh2008} D. Christodoulou proved the formation of a closed trapped surface and eventually a black hole by the focussing of gravitational waves. The incoming radiation has to be large enough in order to form a black hole. This data obeys a specific hierarchy. In particular, the energy $e$ is of the order $e = O(r^{-2} \underline{u}^{-1})$. 

Back to our question: 

In the spacetimes (Bvac) and (B-Tneutrinos), 
it follows that the derivative $\Dlap_3 \hat{\chi}$ takes a well-defined and finite limit at $\mathcal{I}^+$, whereas $\hat{\chi}$ does not. Moreover, note that neither $\rho$ nor $\sigma$ have finite limits at $\mathcal{I}^+$. The corresponding statements are conjectured for spacetimes (Avac) and (A-Tneutrinos).

Consider the EV equations (\ref{EV}) for (Bvac) and (Avac) spacetimes. 
The following will yield important pieces required for the proofs of the theorems \ref{theel***} and \ref{theelmagn***} in (Bvac) spacetimes and provide evidence for the corresponding conjectures \ref{conjtheel***} and \ref{conjtheelmagn***} in (Avac) spacetimes. 

We introduce the notation $\rho_3 : = \Dlap_3  \rho \ + \ \frac{3}{2} tr \underline{\chi} \rho$. 
In the Bianchi equation for $\Dlap_3  \rho$ 
\be  \label{Bianchiturho3}
\Dlap_3  \rho \ + \ \frac{3}{2} tr \underline{\chi} \rho \  =  \ 
- \dlap \underline{\beta} 
 - \frac{1}{2} \hat{\chi} \underline{\alpha} \ + \ 
 ( \epsilon  - \zeta) \underline{\beta} \ + \ 
2 \underline{\xi} \beta 
\ee
we focus on the higher order terms, 
\[
\rho_3 \  =  \ 
 - \underbrace{ \dlap \underline{\beta} }_{= O(r^{-3} \tau_-^{- \frac{1}{2}})}
 - \underbrace{ \frac{1}{2} \hat{\chi} \cdot \underline{\alpha} }_{= O(r^{- \frac{5}{2}} \tau_-^{- \frac{3}{2}})} + \ l.o.t. 
\]
A short computation shows that 
\[
\rho_3 \  =  \ 
 - \underbrace{ \dlap \underline{\beta} }_{= O(r^{-3} \tau_-^{- \frac{1}{2}})} 
- \underbrace{\frac{\partial}{\partial u} (\hat{\chi} \cdot \hat{\underline{\chi}})}_{ = O(r^{- \frac{5}{2}} \tau_-^{- \frac{3}{2}})} + \underbrace{\frac{1}{4} tr \chi  |\hat{\underline{\chi}}|^2 }_{ = O(r^{-3} \tau_-^{-1})} + \ l.o.t. 
\]
Thus it is 
\be \label{rho3*}
\rho_3  + \frac{\partial}{\partial u} (\hat{\chi} \cdot \hat{\underline{\chi}})  \  =  \ 
 - \dlap \underline{\beta} + \frac{1}{4} tr \chi  |\hat{\underline{\chi}}|^2 \ =  \ O(r^{-3} \tau_-^{- \frac{1}{2}}) 
\ee

Similarly, we denote 
$\sigma_3 = \Dlap_3 \sigma + \frac{3}{2} tr \underline{\chi} \sigma$. 
Consider the Bianchi equation 
\[
\sigma_3 \ = \ - \clap \ \ \underline{\beta} - \frac{1}{2} \hat{\chi} \cdot \ ^* \underline{\alpha} + 
\epsilon \ ^* \underline{\beta} - 2 \zeta \ ^* \underline{\beta} - 2 \underline{\xi} \ ^* \beta 
\]
Clearly, the lower order terms decay to zero when approaching null infinity. Then the main part of 
the Bianchi equation reads 
\[
\sigma_3 \ = \ - \clap \ \ \underline{\beta} - \frac{1}{2} \hat{\chi} \cdot \ ^* \underline{\alpha} + l.o.t. 
\]
A short computation yields  
\[
\sigma_3  + \frac{\partial}{\partial u} ( \hat{\chi} \wedge \hat{\underline{\chi}} ) \ = \ - \clap \ \ \underline{\beta} \ = \ 
O(r^{-3} \tau_-^{- \frac{1}{2}})
\]
Recall that for $\sigma$ the same rules apply as for $\rho$ above, and for $\hat{\chi} \wedge \hat{\underline{\chi}}$ the orders of each term are at the level of $\hat{\chi} \cdot \hat{\underline{\chi}}$ above. 

In order to prove the next propositions in this subsection, we will concentrate on equation (\ref{Bianchiturho3}), respectively (\ref{rho3*}), for $\rho_3$, and the corresponding equation for $\sigma_3$. The implications for gravitational waves from the equation for $\rho_3$ will be derived in section \ref{FNI} and from the equation for $\sigma_3$ in section \ref{new}. 

We will multiply (\ref{rho3*}) by $r^3$ and take the limit on $C_u$ as $t \to \infty$. 
First, we see that for each of the leading order terms on the right hand side this limit exists separately. 
The same holds for the leading order terms on the left hand side under the {\bf smallness assumptions} of \cite{lydia1, lydia2}. However, {\bf for large data}, there are additional terms at leading (respectively high) order on the left hand side, that do not tend to a limit at 
$\mathcal{I^+}$, but that cancel. Only the large data case will be considered later. Nevertheless, it is interesting to compare the small with the large data situations, find out what structures are common in both and what are typical structures for large data only. 
The following analyzes these two situations separately.

\subsubsection{A Smallness Assumption} 
\label{small}

In \cite{lydia1, lydia2} it is proven that under the smallness assumptions (\ref{rho3bound}) and (\ref{sigma3bound}) on the initial data in the hypersurface $H_0$ the same inequalities continue to hold on each $H_t$ throughout the spacetimes constructed as global solutions to the EV equations (\ref{EV}). 
A consequence of the proof and the smallness is that $\rho_3 = O(r^{-3} \tau_-^{- \frac{1}{2}})$, 
$\sigma_3 = O(r^{-3} \tau_-^{- \frac{1}{2}})$ 
and thus $\rho_3$ as well as $\sigma_3$ take the corresponding limits at future null infinity. 
Moreover, the structures for $\rho$ (\ref{rhoterms1}) and for $\sigma$ (\ref{sigmaterms1}) emerge under the smallness conditions. 

From this and the structures of the involved terms for small data, it follows that $\frac{\partial}{\partial u} (\hat{\chi} \cdot \hat{\underline{\chi}})$ takes its limit as well. 
In fact, $\frac{\partial}{\partial u} (\hat{\chi} \cdot \hat{\underline{\chi}})$, 
respectively $\int_u \frac{\partial}{\partial u} (\hat{\chi} \cdot \hat{\underline{\chi}}) \ du$,  
behave much better than expected at this point. Namely, for data as in 
\cite{lydia1, lydia2}, we prove in theorem \ref{thhchichi1} that $\int_u \frac{\partial}{\partial u} (\hat{\chi} \cdot \hat{\underline{\chi}}) \ du  = O(r^{-3})$. \\ 

{\bf Bounding $\int_u \frac{\partial}{\partial u} (\hat{\chi} \cdot \hat{\underline{\chi}}) \ du $}: 
\begin{The} \label{thhchichi1}
For spacetimes constructed in \cite{lydia1, lydia2} the following holds: 
\[ 
\int_u \frac{\partial}{\partial u} (\hat{\chi} \cdot \hat{\underline{\chi}}) \ du  = O(r^{-3}) \ . 
\] 
\end{The}
{\itshape Proof of Theorem \ref{thhchichi1}:} 
Consider equation (\ref{rho3*}). Next, we use (\ref{rho3bound}), which is a smallness assumption from \cite{lydia1, lydia2}. 
A consequence of (\ref{rho3bound}) and the proof in \cite{lydia1, lydia2} is that $\rho_3 = O(r^{-3} \tau_-^{- \frac{1}{2}})$. 
On the other hand, equation (\ref{chihat}) implies that $\hat{\chi}$ has a structure as given in (\ref{hchiterms1}). 
Within the spacetimes proven to be stable in \cite{lydia1, lydia2}, the shear $\hat{\underline{\chi}}$ features the structures from (\ref{huchiterms1}). 
It follows that $\int_u \frac{\partial}{\partial u} (\hat{\chi} \cdot \hat{\underline{\chi}}) \ du  = O(r^{-3})$.

\subsubsection{No Assumption, Large Data}
\label{large}

{\bf No smallness assumption, large data:} What happens to these structures if one does not assume any smallness of the data? From now on, the data will be large. The following holds for spacetimes of type (Bvac) and is conjectured for spacetimes of type  (Avac). 
We consider the Bianchi equation (\ref{Bianchiturho3}). Still (\ref{rho3*}) holds, and the leading order terms on the right hand side are not affected. Whereas under the smallness assumption (\ref{rho3bound}), the terms $\rho_3$ as well as $\frac{\partial}{\partial u} (\hat{\chi} \cdot \hat{\underline{\chi}})$ take a well-defined limit at $\mathcal{I}^+$ when multiplied with $r^{3}$, this is not the case anymore if we allow general data. 
For large data many more terms of order $r^{- \frac{5}{2}} \tau_-^{- \frac{3}{2}}$ 
exist in 
$\rho_3$ as well as in $\frac{\partial}{\partial u} (\hat{\chi} \cdot \hat{\underline{\chi}})$ and terms of order 
$r^{- \frac{5}{2}} \tau_-^{-1 - \alpha}$ with $\alpha \geq \frac{1}{2}$. For the latter, note that the known structures for $\frac{\partial}{\partial u} (\hat{\chi} \cdot \hat{\underline{\chi}})$ (see (\ref{hchiterms1}),(\ref{huchiterms1})) require $\alpha \geq \frac{1}{2}$ and enforce this via equation (\ref{rho3*}) on $\rho_3$. 
The corresponding statement holds for $\frac{\partial}{\partial u} (\hat{\chi} \wedge \hat{\underline{\chi}})$ and $\sigma_3$ via the Bianchi equation for the latter. 

Summarizing the above, we find that for large data the following quantities enjoy the structures 
\bea
\rho & = & [ r^{- \frac{5}{2}} ] +  \{ r^{- \frac{5}{2}}  \tau_-^{- \alpha} \} + \{ r^{-3}  \tau_-^{+ \frac{1}{2}}  \} + \{ r^{-3}  \tau_-^{+ \beta}  \} 
+ \{ r^{- 3} \}  \label{Allgrhoterms1} 
+  O(r^{-3} \omega^{-\gamma})  
\eea
and 
\bea
\sigma & = & [ r^{- \frac{5}{2}} ] +  \{ r^{- \frac{5}{2}}  \tau_-^{- \alpha} \} + \{ r^{-3}  \tau_-^{+ \frac{1}{2}}  \} + \{ r^{-3}  \tau_-^{+ \beta}  \} 
+ \{ r^{- 3} \} \label{Allgsigmaterms1} 
+  O(r^{-3} \omega^{-\gamma}) 
\eea
with $\omega$ denoting $r$ or $\tau_-$ and $\gamma > 0$, $0 < \beta < \frac{1}{2}$, $\alpha \geq \frac{1}{2}$, as well as 
\bea
 (\hat{\chi} \cdot \hat{\underline{\chi}}) & = &  \{ r^{- \frac{5}{2}}  \tau_-^{- \alpha} \}  + \{ r^{-3} \} +   O(r^{-3} \omega^{-\gamma})  \label{Allgchiuchiterms1} \\ 
 (\hat{\chi} \wedge \hat{\underline{\chi}}) & = &  \{ r^{- \frac{5}{2}}  \tau_-^{- \alpha} \}  + \{ r^{-3} \} +   O(r^{-3} \omega^{-\gamma})  \label{Allgchiuchiterms2}
\eea
Note that the second term on the right hand side of (\ref{Allgchiuchiterms1}) is dynamical, thus depends on $u$. The same holds for the second term on the right hand side of (\ref{Allgchiuchiterms2}). 

From the specific structures given above of the involved derivatives of $\rho, \sigma$ and $\hat{\chi}, \hat{\underline{\chi}}$ we derive 
\beas
\rho_3 - \{ r^{- \frac{5}{2}}  \tau_-^{-1 - \alpha} \}  & =  & \underbrace{\{ r^{-3} \tau_-^{- \frac{1}{2}} \}}_{:= \rho_{\frac{1}{2}}}
+ \underbrace{\{ r^{-3}  \tau_-^{ -1 + \beta }  \} }_{:= \rho_{\beta}}
+ l.o.t. 
\ = \  \rho_{\frac{1}{2}} + \rho_{\beta} + l.o.t. \\ 
& = & 
\rho_{\frac{1}{2}} + \rho_{\beta} +  \underbrace{\rho'_f + O(r^{-3} \omega^{-1 -\gamma})}_{l.o.t.}   \\ 
\sigma_3 - \{ r^{- \frac{5}{2}}  \tau_-^{-1 - \alpha} \}  & =  & \underbrace{\{ r^{-3} \tau_-^{- \frac{1}{2}} \}}_{:= \sigma_{\frac{1}{2}}}
+ \underbrace{\{ r^{-3}  \tau_-^{ -1 + \beta }  \} }_{:= \sigma_{\beta}}
+ l.o.t. 
\ = \ 
\sigma_{\frac{1}{2}} + \sigma_{\beta} + l.o.t.  \\ 
& = & 
\sigma_{\frac{1}{2}} + \sigma_{\beta} +  \underbrace{\sigma'_f + O(r^{-3} \omega^{-1 -\gamma})}_{l.o.t.}   \\
\frac{\partial}{\partial u} (\hat{\chi} \cdot \hat{\underline{\chi}}) - \{ r^{- \frac{5}{2}}  \tau_-^{-1 - \alpha} \}  & = & l.o.t. 
\ = \ 
\underbrace{d' + O(r^{-3} \omega^{-1 -\gamma})}_{l.o.t.}  \\ 
\frac{\partial}{\partial u} (\hat{\chi} \wedge \hat{\underline{\chi}}) - \{ r^{- \frac{5}{2}}  \tau_-^{-1 - \alpha} \}  & = & l.o.t. 
\ = \ 
\underbrace{g' + O(r^{-3} \omega^{-1 -\gamma})}_{l.o.t.}  
\eeas
whereby we introduce the new notation $\rho_{\frac{1}{2}}, \rho_{\beta}, \sigma_{\frac{1}{2}}, \sigma_{\beta}$ and 
$\rho'_f, \sigma'_f, d', g'$. 

From these we can immediately conclude proposition \ref{rho_3etal****} and provide supporting evidence for conjecture \ref{urho_3etal****}.

\begin{prop} \label{rho_3etal****}
In (Bvac) spacetimes the following holds: 
\beas
\int_u \Big( \rho_3 - \{ r^{- \frac{5}{2}}  \tau_-^{-1 - \alpha} \} \Big)  \ du  & =  & \int_u \rho_{\frac{1}{2}} \ du \ + \ \int_u \rho_{\beta} \ du \ + \{ r^{- 3} \} 
+  O(r^{-3} \omega^{-\gamma})   \\ 
\int_u \Big( \sigma_3 - \{ r^{- \frac{5}{2}}  \tau_-^{-1 - \alpha} \} \Big)  \ du  & =  &  \int_u  \sigma_{\frac{1}{2}}  \ du \ + \   \int_u \sigma_{\beta}  \ du \ + \  \{ r^{- 3} \} 
+  O(r^{-3} \omega^{-\gamma})  \\ 
\int_u \Big( \frac{\partial}{\partial u} (\hat{\chi} \cdot \hat{\underline{\chi}}) \ - \{ r^{- \frac{5}{2}}  \tau_-^{-1 - \alpha} \} \Big)  \ du & = & 
\{ r^{- 3} \} 
+  O(r^{-3} \omega^{-\gamma})  \\ 
\int_u \Big( \frac{\partial}{\partial u} (\hat{\chi} \wedge \hat{\underline{\chi}}) \ - \{ r^{- \frac{5}{2}}  \tau_-^{-1 - \alpha} \} \Big)  \ du & = & 
\{ r^{- 3} \} 
+  O(r^{-3} \omega^{-\gamma})  
\eeas 
where $\omega$ denotes $r$ or $\tau_-$ and $\gamma > 0$, $0 < \beta < \frac{1}{2}$, $\alpha \geq \frac{1}{2}$. 
In particular, it is 
\beas
\int_u \rho'_f \ du & = &  \{ r^{- 3} \} \ , \ \ \int_u \sigma'_f \ du  =   \{ r^{- 3} \} \ , \ \  
 \int_u d' \ du  =   \{ r^{- 3} \} \ , \ \  \int_u g' \ du  =   \{ r^{- 3} \} \ . 
\eeas
\end{prop}

\begin{conj} \label{urho_3etal****}
The statements of proposition \ref{rho_3etal****} hold for (Avac) spacetimes correspondingly. 
\end{conj}

{\itshape Notation:} There may be cancellations in 
$\rho'_f + d'$ and in $\sigma'_f + g'$. We denote the remaining sum as $\rho_f + d$, respectively $\sigma_f + g$, where $\rho_f$ denotes the corresponding terms from $\rho_3$, and $\sigma_f$ denotes the corresponding terms from $\sigma_3$, correspondingly $d$ relates to $\frac{\partial}{\partial u} (\hat{\chi} \cdot \hat{\underline{\chi}})$ and $g$ to 
$\frac{\partial}{\partial u} (\hat{\chi} \wedge \hat{\underline{\chi}})$.

As a consequence of equation (\ref{rho3*}) all the higher order terms on the left hand side of (\ref{rho3*}), that do not attain a well-defined limit at $\mathcal{I}^+$, cancel. In particular, we find 
\be \label{slandi}
\rho_3  + \frac{\partial}{\partial u} (\hat{\chi} \cdot \hat{\underline{\chi}}) = 
\{ r^{-3} \tau_-^{- \frac{1}{2}} \} + 
\{ r^{-3}  \tau_-^{ -1 + \beta }  \} 
+ l.o.t. 
\ = \ \rho_{\frac{1}{2}} + \rho_{\beta} + \rho_f + d + O(r^{-3} \omega^{-1 -\gamma}) 
\ee
with $0 < \beta < \frac{1}{2}$, where there are more structures in ``l.o.t." with both contributions $\rho_f$ from $\rho_3$ as well as $d$ from 
$\frac{\partial}{\partial u} (\hat{\chi} \cdot \hat{\underline{\chi}})$, which we shall investigate in the subsection \ref{FNI}. 

Using proposition \ref{rho_3etal****}, it follows that 
on the right hand side of (\ref{slandi}), all the quantities of the resulting higher orders, namely $\rho_{\frac{1}{2}}$ and $\rho_{\beta}$, 
(and that are not absorbed in ``l.o.t.") stem from $\rho_3$. This will have direct consequences in formula (\ref{limitstructures}). 

In a similar way, we conclude from 
the Bianchi equation for $\sigma_3$ above that 
\be \label{slandi2}
\sigma_3  + \frac{\partial}{\partial u} ( \hat{\chi} \wedge \hat{\underline{\chi}} ) = 
\{ r^{-3} \tau_-^{- \frac{1}{2}} \} + 
\{ r^{-3}  \tau_-^{ -1 + \beta }  \} 
+ l.o.t. 
\ = \ \sigma_{\frac{1}{2}} + \sigma_{\beta} + \sigma_f + g + O(r^{-3} \omega^{-1 -\gamma}) 
\ee
To investigate the deeper structures of the this equation and its limit at $\mathcal{I}^+$, including the lower order terms, will be the subject of subsection \ref{new}. 

Again, we use proposition \ref{rho_3etal****}, to conclude that 
on the right hand side of (\ref{slandi2}), all the quantities of the resulting higher orders, namely $\sigma_{\frac{1}{2}}$ and $\sigma_{\beta}$, 
(and that are not absorbed in ``l.o.t.") stem from $\sigma_3$. This will have direct consequences in formula \ref{limitstructuresdawn}.

\begin{prop} \label{Rholimits1}
In (Bvac) spacetimes, the following holds for the domain of dependence of the complement of a sufficiently large compact subset of the initial hypersurface. 
The quantities 
$r^3 \rho_{\frac{1}{2}}, r^3 \rho_{\beta}, r^3 \sigma_{\frac{1}{2}}, r^3 \sigma_{\beta}$ have limits on any null hypersurface $C_u$ as $t \to \infty$. Namely, 
\begin{eqnarray*}
 \lim_{C_u, t \to \infty} (r^3 \rho_{\frac{1}{2}} ) = \mathcal{R}_{\frac{1}{2}} (u, \cdot) \ \ , 
\ \ \ \ \ \ \ \ \ \ \ \ \lim_{C_u, t \to \infty} (r^3 \rho_{\beta} ) = \mathcal{R}_{\beta} (u, \cdot) \\ 
\lim_{C_u, t \to \infty} (r^3 \sigma_{\frac{1}{2}} ) = \mathcal{S}_{\frac{1}{2}} (u, \cdot) \ \ , 
\ \ \ \ \ \ \ \ \ \ \ \ \lim_{C_u, t \to \infty} (r^3 \sigma_{\beta} ) = \mathcal{S}_{\beta} (u, \cdot) 
\end{eqnarray*}
where $\mathcal{R}_{\frac{1}{2}} (u, \cdot), \mathcal{R}_{\beta} (u, \cdot), \mathcal{S}_{\frac{1}{2}} (u, \cdot), \mathcal{S}_{\beta} (u, \cdot)$ are well defined on $\real \times S^2$, and satisfy 
\begin{eqnarray*}
\left| \mathcal{R}_{\frac{1}{2}} \left( u,\cdot \right) \right| &\leq &C\left( 1+\left| u\right|
\right) ^{-1/2}\, \, \ \ \ \ \ \ \ \ \ \ \ \ \left| \mathcal{R}_{\beta} \left( u,\cdot
\right) \right| \leq C\left( 1+\left| u\right| \right) ^{-1 + \beta}  \\ 
\left| \mathcal{S}_{\frac{1}{2}} \left( u,\cdot \right) \right| &\leq &C\left( 1+\left| u\right|
\right) ^{-1/2}\, \, \ \ \ \ \ \ \ \ \ \ \ \ \left| \mathcal{S}_{\beta} \left( u,\cdot
\right) \right| \leq C\left( 1+\left| u\right| \right) ^{-1 + \beta} 
\end{eqnarray*}
for $0 < \beta < \frac{1}{2}$. 

Moreover, $r^3 \rho_f, \ r^3 \sigma_f, \ r^3 d, \ r^3 g$ attain well-defined limits on any null hypersurface $C_u$ as $t \to \infty$. These limits depend on $u$. 
\end{prop}
We denote by $| \ |$ the pointwise norms on $S^2$ with respect to the standard metric. 

\begin{conj} \label{uRholimits1}
The statements of proposition \ref{Rholimits1} hold for 
(Avac) spacetimes correspondingly. 
\end{conj}

{\itshape Proof of Proposition \ref{Rholimits1}:} The proof of proposition \ref{Rholimits1} also provides evidence for conjecture  \ref{uRholimits1}. 

For (Bvac) spacetimes the 
existence of a complete domain of dependence of the complement of a sufficiently large compact subset of the initial hypersurface is guaranteed by the main theorem in \cite{lydia1, lydia2} together with the content of the last part of section \ref{slgrm}. 

Then proposition \ref{Rholimits1} follows immediately from the calculations leading up to proposition \ref{rho_3etal****}. \\

Now, consider the limit at $\mathcal{I}^+$ of the left hand side of (\ref{rho3*}). 
It follows from (\ref{slandi}) and proposition \ref{Rholimits1} that 
\be \label{Rholimitsup2}
\lim_{C_u, t \to \infty} r^3 \big( \rho_3  + \frac{\partial}{\partial u} (\hat{\chi} \cdot \hat{\underline{\chi}})  \big) = 
\mathcal{R}_{\frac{1}{2}} (u, \cdot) + \mathcal{R}_{\beta} (u, \cdot) + l.o.t. 
\ee

The integral with respect to $u$ of the leading order term emerges as $O(r^{-3} \tau_-^{+ \frac{1}{2}})$. This and related questions will be investigated next in subsection \ref{FNI}. 

Similarly, it follows from (\ref{slandi2}) and proposition \ref{Rholimits1} that 
\be \label{Sigmalimitsup2}
\lim_{C_u, t \to \infty} r^3 \big( \sigma_3  + \frac{\partial}{\partial u} (\hat{\chi} \wedge \hat{\underline{\chi}})  \big) = 
\mathcal{S}_{\frac{1}{2}} (u, \cdot) + \mathcal{S}_{\beta} (u, \cdot) + l.o.t. 
\ee 

Again, the integral with respect to $u$ of the leading order term emerges as $O(r^{-3} \tau_-^{+ \frac{1}{2}})$. This and related questions will be investigated in subsection \ref{new}. \\

{\bf Remarks:} \\
$\bullet$ In $\int_u \frac{\partial}{\partial u} (\hat{\chi} \cdot \hat{\underline{\chi}}) \ du$ as well as in $\int_u \frac{\partial}{\partial u} (\hat{\chi} \wedge \hat{\underline{\chi}}) \ du$ 
the terms of order $O(r^{-3})$ in general are not independent of $u$. \\ 
$\bullet$ Note that for AF systems with fall-off towards infinity of $O(r^{-1})$, the product $(\hat{\chi} \cdot \hat{\underline{\chi}})$, respectively $(\hat{\chi} \wedge \hat{\underline{\chi}})$, decays in $|u|$. Namely, it is 
$(\hat{\chi} \cdot \hat{\underline{\chi}}) = O(r^{-3} \tau_-^{- \alpha})$, respectively $(\hat{\chi} \wedge \hat{\underline{\chi}}) = O(r^{-3} \tau_-^{- \alpha})$, 
with $\alpha > 0$.

\subsection{Future Null Infinity and Electric Memory}
\label{FNI}

Recall that we consider {\bf large data}. Let us investigate (Bvac) and (Avac) spacetimes. 
The following arguments and computations will prove theorem \ref{theel***} and provide evidence for conjecture \ref{conjtheel***}. 
Parts of the emerging structures and results will also be used to establish theorem \ref{theelmagn***} and conjecture \ref{conjtheelmagn***}. 

Above we found that 
the left hand side of (\ref{rho3*}) is of order $O(r^{-3} \tau_-^{- \frac{1}{2}})$. 
We introduce the following notation for the corresponding limit of the left hand side of (\ref{rho3*}): 
\bea
\mathcal{P}_{3} & : =  & \lim_{C_u, t \to \infty} r^3 \big{(} \rho_3  + \frac{\partial}{\partial u} (\hat{\chi} \cdot \hat{\underline{\chi}}) \big{)}  \label{defrho3*limit**1} \\ 
\mathcal{P} & : =  & \int_u \mathcal{P}_3 \ du \label{defP3*2limit**1} 
\eea
Note that $\mathcal{P}$ is defined on $S^2 \times \real$ up to an additive function $C_{\mathcal{P}}$ on $S^2$ (thus the latter is independent of $u$). Later, when taking the integral 
$\int_{- \infty}^{+ \infty} \mathcal{P}_3 \ du$, the term $C_{\mathcal{P}}$ will cancel.

Taking the limit of $\big{(}$$r^3$ (\ref{rho3*})$\big{)}$ on $C_u$ as $t \to \infty$, 
each term on the right hand side takes a well-defined limit. 
This yields 
\be \label{Lrho3*}
\mathcal{P}_3  \ = \ 
- \dlap \underline{B} + 2 | \Xi |^2 
\ee

Consider (\ref{Rholimitsup2}) in connection with proposition \ref{Rholimits1}, from which it follows that 
\be \label{P3brook1}
\mathcal{P}_3  \ = \ \mathcal{R}_{\frac{1}{2}} (u, \cdot) + \mathcal{R}_{\beta} (u, \cdot) + l.o.t. 
\ee
For the next step, we shall take into account the deeper structures of the lower order terms in (\ref{P3brook1}) as well. 

Now, we look at the details for $\mathcal{P}$ in (\ref{defP3*2limit**1}). 

\begin{prop} \label{Pres}
For (Bvac) spacetimes, $\mathcal{P}$ has the following structure for $0 < \beta < \frac{1}{2}$ and $\gamma > 0$, 
\bea 
\mathcal{P} \ & = & \ \underbrace{\{ \tau_-^{+ \frac{1}{2}} \} \ + \ \{  \tau_-^{\beta} \}}_{ = \mathcal{P}_{\rho_1}}  \ + \ 
\underbrace{\{ \mathcal{F} (u, \cdot) \}}_{= \mathcal{P}_{\rho_2 } - \frac{1}{2} D} \ + \ \{  \tau_-^{-  \gamma} \} \  + \ C_{\mathcal{P}}   \label{limitstructures}
\eea
where $\mathcal{F} (u, \cdot) \leq C$. Again $\{ \cdot \}$ means terms of order given inside the brackets. And $C_{\mathcal{P}}$ is an 
additive function on $S^2$ introduced above. 
\end{prop}
{\itshape Proof of Proposition \ref{Pres}:} 
Proposition \ref{Rholimits1} yields directly that 
\be \label{Rho1234}
\int_u \mathcal{R}_{\frac{1}{2}} (u, \cdot) \ du \ = \ \{ \tau_-^{+ \frac{1}{2}} \} 
\ee
and 
\be \label{Rho1234beta}
\int_u \mathcal{R}_{\beta} (u, \cdot)  \ du \ = \  \{  \tau_-^{\beta} \} 
\ee
From (\ref{P3brook1}) together with proposition \ref{Rholimits1} it follows that 
\[
\mathcal{P} \ = \ \int_u \mathcal{R}_{\frac{1}{2}} (u, \cdot) \ du \ + \ \int_u \mathcal{R}_{\beta} (u, \cdot)  \ du \ + \ l.o.t. 
\]
Thus, the first term on the right hand side of (\ref{limitstructures}) is given by (\ref{Rho1234}) and the second term on the right hand side of (\ref{limitstructures}) is given by (\ref{Rho1234beta}). As these quantities are rooted in $\rho_3$, we denote 
$( \{ \tau_-^{+ \frac{1}{2}} \} \ + \ \{  \tau_-^{\beta} \} )$ by $\mathcal{P}_{\rho_1}$. Thus, the terms of order $O( \tau_-^{\alpha})$ with $0 < \alpha \leq \frac{1}{2}$ originate from the integral of the corresponding limits of the $\rho_3$ part.

In order to analyze the third term on the right hand side of (\ref{limitstructures}), we recall proposition \ref{rho_3etal****}. 
Taking the corresponding limits in (\ref{defrho3*limit**1}) and (\ref{defP3*2limit**1}) it turns out that $\{ \mathcal{F} (u, \cdot) \}$ depends on $u$ but does not decay in $u$. 
That is, 
$\mathcal{F} (u, \cdot)$ comprises the corresponding components of the integral in $u$ originating from the limits $\lim_{C_u, t \to \infty} (r^3 \rho_f)$, respectively $\lim_{C_u, t \to \infty} (r^3 d)$ 
of the terms 
$\rho_f$, respectively $d$ in (\ref{slandi}). 
Therefore, $\mathcal{F} (u, \cdot)$ has pieces that are sourced by $\rho_3$ and pieces that are sourced by $\frac{\partial}{\partial u} (\hat{\chi} \cdot \hat{\underline{\chi}})$, we denote the former by $\mathcal{P}_{\rho_2 }$ and the latter by $- \frac{1}{2} D$.

Similarly, for the fourth term on the right hand side of (\ref{limitstructures}), recall proposition \ref{rho_3etal****} and 
take the corresponding limits in (\ref{defrho3*limit**1}) and (\ref{defP3*2limit**1}). We find that $\{  \tau_-^{-  \gamma} \}$ is sourced by terms that are rooted in $\rho_3$ as well as in $\frac{\partial}{\partial u} (\hat{\chi} \cdot \hat{\underline{\chi}})$. As $\{  \tau_-^{-  \gamma} \}$ decay in $u$, they will not contribute to the effects to be studied. 

Finally, $C_{\mathcal{P}}$ in (\ref{limitstructures}) is the said constant that will cancel later. 

This proves proposition \ref{Pres}. 

Using the structures of proposition \ref{Pres} and its proof, 
we write 
\be \label{limitstructures2}
\mathcal{P} (u, \cdot) \ = \ \mathcal{P}_{\rho_1} (u, \cdot) \ + \ \mathcal{P}_{\rho_2 } (u, \cdot) \ - \frac{1}{2}  \ D (u, \cdot) \  + \ C_{\mathcal{P}}  \ + \ l.o.t. 
\ee
We emphasize that $\mathcal{P}_{\rho_2 }$ as well as $D$ depend on $u$.

Next, we define 
\bea
Chi_3 & : =  & \lim_{C_u, t \to \infty} \big{(}  r^2 \frac{\partial}{\partial u}  \hat{\chi} \label{defChi3*1} \big{)} \\ 
Chi & : =  & \int_u Chi_3 \ du  \label{defChi3*2}
\eea
For these spacetimes it follows from (\ref{dchibund1}) that (\ref{Ldchibund1}) holds, namely
\[
\underline{B} = - 2 \dlap \Xi 
\]
and from (\ref{chihat}) that 
\be \label{Lchihat}
Chi_3 = - \Xi 
\ee
Using (\ref{Ldchibund1}) and (\ref{Lchihat}) in (\ref{Lrho3*}) gives 
\be \label{LCrho3*}
\mathcal{P}_3  \ = \ 
- 2 \dlap \dlap Chi_3 + 2 | \Xi |^2 
\ee
Integrating (\ref{LCrho3*}) with respect to $u$ gives 
\be \label{supergold****}
(\mathcal{P}^- - \mathcal{P}^+) - \int_{- \infty}^{+ \infty} | \Xi |^2 \ du 
  \ = \ 
\dlap \dlap (Chi^- - Chi^+) 
\ee
Recall (\ref{limitstructures}) and (\ref{limitstructures2}) 
and write 
\bea 
& &
 (\mathcal{P}_{\rho_1 }^- - \mathcal{P}_{\rho_1 }^+)
\ + \ (\mathcal{P}_{\rho_2 }^- - \mathcal{P}_{\rho_2 }^+) 
\ - \  \frac{1}{2} (D^- - D^+)  \nonumber \\ 
& & 
- \int_{- \infty}^{+ \infty} | \Xi |^2 \ du   \nonumber \\ 
& & 
\ = \ 
\dlap \dlap (Chi^- - Chi^+) \label{supergold}
\eea
First we see that the last term on the left hand side 
$(- \int_{- \infty}^{+ \infty} | \Xi |^2 \ du)$ is finite, in fact it is borderline within the solutions of \cite{lydia1, lydia2}. 
Thus, it is finite for (Bvac) spacetimes. However, it may not be bounded for general (Avac) spacetimes. 
We see that $(\mathcal{P}^- - \mathcal{P}^+)$, respectively $\dlap \dlap (Chi^- - Chi^+)$ are infinite for (Bvac) as well as (Avac) spacetimes. 
To see this, we recall the concept of limits at null infinity $\mathcal{I^+}$ introduced in section \ref{limits12}. 
Now, fix a point on the sphere $S^2$ at fixed $u_0$ and consider $\mathcal{P}(u_0)$. Next, take $\mathcal{P}(u)$ at the corresponding point for some value of $u \neq u_0$. Keep $u_0$ fixed and let $u$ tend to $+ \infty$, respectively to $- \infty$. 
Then the difference $\mathcal{P}(u) - \mathcal{P}(u_0)$ is no longer finite, but it grows with $|u|^{+ \frac{1}{2}}$ for (Bvac) as well as (Avac) spacetimes. A corresponding argument holds for $Chi(u) - Chi(u_0)$. 
The extra term $(D^- - D^+)$ is finite. 
Note that for any asymptotically flat system of decay $O(r^{-1})$ towards infinity, ${D}$ 
decays in $|u|$ for large $|u|$ fast enough, and $\hat{\chi}$ as well as $\hat{\underline{\chi}}$ take limits at $\mathcal{I}^+$ such that 
the product of their limits vanishes as $|u| \to \infty$. Therefore, in those systems one has $D^- = D^+ = 0$. 

It follows that there exists a function $\Phi$ such that 
\bea
 \dlap (Chi^- - Chi^+) & = & \nlap \Phi + X  \label{Phi1a} \\ 
\dlap \dlap (Chi^- - Chi^+) & = & \slap \Phi  \nonumber \\ 
& = & (\mathcal{P} - \bar{\mathcal{P}})^- - (\mathcal{P} - \bar{\mathcal{P}})^+ \nonumber \\ 
& & - 2 (F - \bar{F})  \label{Phi2a} 
\eea
where $X = \nlap^{\perp} \Psi$ for a function $\Psi$ whose Laplacian $\slap \Psi = \clap \ \  \dlap (Chi^- - Chi^+)$. However, in order to reveal the new structures more clearly, in this section, we put the contribution from $\Psi$ to zero. In section \ref{new}, we treat the most general case with $\nlap^{\perp} \Psi \neq 0$ to obtain the complete general result and the full set of equations 
(\ref{Psi1})-(\ref{Phi33}). 
Note that D. Christodoulou showed in \cite{chrmemory} that for spacetimes as studied by Christodoulou and Klainerman in \cite{sta} there is no contribution from the $\clap \ \ $ equation. The same holds for any AF system that decays towards infinity like $O(r^{-1})$ as shown by the present author in \cite{lydia4}. Thus, the works \cite{chrmemory} and \cite{lydia4} prove that there is no magnetic memory for those systems. 
We will see in equations (\ref{Psi1})-(\ref{Psi2}) in the system (\ref{Psi1})-(\ref{Phi33}) that the magnetic memory occurs naturally 
in the (Bvac) and (Avac) spacetimes.

In our present setting, using the above structures, we have 
\bea
 \dlap (Chi^- - Chi^+) & = & \nlap \Phi   \label{Phi1} \\ 
\dlap \dlap (Chi^- - Chi^+) & = & \slap \Phi  \nonumber \\ 
& = & (\mathcal{P}_{\rho_1 } - \bar{\mathcal{P}}_{\rho_1 })^- - (\mathcal{P}_{\rho_1 } - \bar{\mathcal{P}}_{\rho_1 })^+ \nonumber \\ 
& & (\mathcal{P}_{\rho_2 } - \bar{\mathcal{P}}_{\rho_2 })^- - (\mathcal{P}_{\rho_2 } - \bar{\mathcal{P}}_{\rho_2 })^+ \nonumber \\ 
& & - 2 (F - \bar{F})  \label{Phi2} \\ 
& & - \frac{1}{2} (D - \bar{D})^- +  \frac{1}{2} (D - \bar{D})^+ \nonumber
\eea 
Hodge theory provides the solution on $S^2$. 

The resulting difference $Chi^- - Chi^+$ is related to the permanent change of the distance of nearby geodesics, manifesting itself in a permanent displacement of test masses in a gravitational wave detector like LIGO. In particular, $Chi^- - Chi^+$ multiplied by a factor including the initial distance of the test masses encodes the displacement of test masses given by the ordinary and the null memory effects. 
See details in \cite{chrmemory} and \cite{lydia4}. 

The memory given by equations (\ref{Phi1})-(\ref{Phi2}) (that is (\ref{Phi1a})-(\ref{Phi2a})) is of {\bf electric} parity. It consists of the finite null memory generated by the radiated energy $F$ (finite for (Bvac), unbounded for (Avac) spacetimes) and the infinite ordinary memory due to the part 
$(\mathcal{P} - \bar{\mathcal{P}})^- - (\mathcal{P} - \bar{\mathcal{P}})^+$, in particular $ (\mathcal{P}_{\rho_1 } (u) - \bar{\mathcal{P}}_{\rho_1 } (u)) - (\mathcal{P}_{\rho_1 } (u_0) - \bar{\mathcal{P}}_{\rho_1 } (u_0))$ for fixed $u_0$ and $u$ tending to $+ \infty$, respectively to $- \infty$, is growing like $|u|^{+ \frac{1}{2}}$. This memory is due to the gravitational waves radiating. 
Finite memory is added by $(\mathcal{P}_{\rho_2 } - \bar{\mathcal{P}}_{\rho_2 })^- - (\mathcal{P}_{\rho_2 } - \bar{\mathcal{P}}_{\rho_2 })^+$ as well as 
$- \frac{1}{2}(D - \bar{D})^- +  \frac{1}{2} (D - \bar{D})^+$. 

This {\bf proves} the following theorem. 

\begin{The} \label{theel***}
The following holds for (Bvac) spacetimes. 

If $\nlap^{\perp} \Psi \equiv 0$, then 
$(Chi^- - Chi^+)$ is determined by equation (\ref{Phi1}) on $S^2$ where $\Phi$ is the solution with vanishing mean of (\ref{Phi2}). 

\end{The}

The above gives the details on the finer structures of the divergent memory derived in \cite{lydia4}.

\begin{conj} \label{conjtheel***}
The statement of theorem \ref{theel***} holds correspondingly for (Avac) spacetimes. 
\end{conj}

\subsection{Future Null Infinity and Magnetic Memory}
\label{new}

In this section, we treat the most general case with $\nlap^{\perp} \Psi \neq 0$. 

We observe, that in these general spacetimes of very slow decay of the data towards infinity, the quantities $(\hat{\chi} \cdot \hat{\underline{\chi}})$ as well as $(\hat{\chi} \wedge \hat{\underline{\chi}})$ play a more dominant role than for systems falling off like $O(r^{-1})$. In particular, due to the non-vanishing of the corresponding limits for large $|u|$, 
they impact $\mathcal{I}^+$. 

In section \ref{FNI} we derived the electric memory with its various characteristics. 
How about magnetic memory? We know that the latter does not occur in AF systems falling off like $O(r^{-1})$, \cite{chrmemory}, \cite{lydia4}. 
The answer for more general situations is very different. 

In what follows, we are going to prove that magnetic memory exists in spacetimes (Bvac) and is conjectured for (Avac) (shown in this section \ref{new}) and we are going to prove that magnetic memory exists in spacetimes 
(B-Tneutrinos) and is conjectured for (A-Tneutrinos) (shown in section \ref{neutrinos}). Thereby we are going to derive the precise structures. This magnetic memory grows with $\sqrt{|u|}$. It arises naturally in the realm of the Einstein vacuum equations as well as for the Einstein-null-fluid equations describing neutrino radiation. It consists of various parts with interesting characteristics. 
As a most striking feature, (A-Tneutrinos) solutions exhibit an additional contribution due to a curl term of the stress-energy tensor. 

Next, we are going to derive the magnetic memory effects for (Bvac) spacetimes and provide evidence for (Avac) spacetimes. 
In particular, the following arguments and computations will prove theorem \ref{theelmagn***} and provide evidence for conjecture \ref{conjtheelmagn***}. 

Recall from above that 
$\sigma_3 = \Dlap_3 \sigma + \frac{3}{2} tr \underline{\chi} \sigma$. 
Consider the Bianchi equation 
\[
\sigma_3 \ = \ - \clap \ \ \underline{\beta} - \frac{1}{2} \hat{\chi} \cdot \ ^* \underline{\alpha} + 
\epsilon \ ^* \underline{\beta} - 2 \zeta \ ^* \underline{\beta} - 2 \underline{\xi} \ ^* \beta 
\]
As we saw previously, the lower order terms decay to zero when approaching null infinity. Then the main part of 
the Bianchi equation reads 
\be \label{s313}
\sigma_3 \ = \ - \clap \ \ \underline{\beta} - \frac{1}{2} \hat{\chi} \cdot \ ^* \underline{\alpha} + l.o.t. 
\ee
and a short computation yields  
\be \label{s314}
\sigma_3  + \frac{\partial}{\partial u} ( \hat{\chi} \wedge \hat{\underline{\chi}} ) \ = \ - \clap \ \ \underline{\beta} \ = \ 
O(r^{-3} \tau_-^{- \frac{1}{2}})
\ee
Recall the structures derived in propositions \ref{rho_3etal****} and \ref{Rholimits1} for (Bvac) spacetimes and the conjectures \ref{urho_3etal****} and \ref{uRholimits1} for (Avac) spacetimes. 
The emerging behavior in the present settings are in contrast to those in spacetimes with faster decay, namely systems with 
$g_{\mu \nu} - \eta_{\mu \nu} = O(r^{-1})$. 

The following new results hold for (Bvac) spacetimes and are conjectured for (Avac) spacetimes:

The highest order terms on the left hand side of (\ref{s314}) cancel, and the remaining terms are of order $O(r^{-3} \tau_-^{- \frac{1}{2}})$. 
Multiply the left hand side of (\ref{s314}) by $r^3$ and and take the limit on each $C_u$ for $t \to \infty$ denoting this limit by 
$\mathcal{Q}_3$. Then introduce $\mathcal{Q}$ to be its integral with respect to $u$. 
\bea 
\mathcal{Q}_3 & : = & \lim_{C_u, t \to \infty} r^3 \big{(} \sigma_3  + \frac{\partial}{\partial u} ( \hat{\chi} \wedge \hat{\underline{\chi}} ) \big{)}   \label{defQ3****2} \\  
\mathcal{Q} &  : =  &  \int_u \mathcal{Q}_3 \ du   \label{defQ3*2} 
\eea
$\mathcal{Q}$ is defined on $S^2 \times \real$ up to an additive function $C_{\mathcal{Q}}$ on $S^2$ (thus the latter is independent of $u$). Later, when taking the integral 
$\int_{- \infty}^{+ \infty} \mathcal{Q}_3 \ du$, the term $C_{\mathcal{Q}}$ will cancel. 

Take the limit of ($r^3$ (\ref{s314})) on $C_u$ as $t \to \infty$ to obtain 
\be \label{Q3limit1234}
\mathcal{Q}_3 \ = \ - \clap \ \ \underline{B} 
\ee

Consider (\ref{Sigmalimitsup2}) together with proposition \ref{Rholimits1} to conclude 
\be \label{Q3Sigma1234*}
\mathcal{Q}_3 \ = \ \mathcal{S}_{\frac{1}{2}} (u, \cdot) + \mathcal{S}_{\beta} (u, \cdot) + l.o.t. 
\ee 
In a first step, we singled out the leading structures in $\mathcal{S}_{\frac{1}{2}} (u, \cdot)$ and $\mathcal{S}_{\beta} (u, \cdot)$ on the left hand side of (\ref{Q3Sigma1234*}), whereas in the next step we will investigate the lower order terms as well. 

Next, we shall investigate the deeper structures of $\mathcal{Q}$ in (\ref{defQ3*2}).

\begin{prop} \label{Qres}
For (Bvac) spacetimes, $\mathcal{Q}$ has the following structure for $0 < \beta < \frac{1}{2}$ and $\gamma > 0$, 
\bea 
\mathcal{Q} \ & = & \ \underbrace{\{ \tau_-^{+ \frac{1}{2}} \} \ + \ \{  \tau_-^{\beta} \}}_{ = \mathcal{Q}_{\sigma_1}}  \ + \ 
\underbrace{\{ \mathcal{F} (u, \cdot) \}}_{= \mathcal{Q}_{\sigma_2 } - \frac{1}{2}  G} \ + \ \{  \tau_-^{-  \gamma} \} \ + \ C_{\mathcal{Q}} 
\label{limitstructuresdawn}
\eea
where $\mathcal{F} (u, \cdot) \leq C$. Again $\{ \cdot \}$ means terms of order given inside the brackets. And $C_{\mathcal{Q}}$ is an 
additive function on $S^2$ not depending on $u$. 
\end{prop}
{\itshape Proof of Proposition \ref{Qres}:} 
Proposition \ref{Rholimits1} yields directly that 
\be \label{Sigma1234}
\int_u \mathcal{S}_{\frac{1}{2}} (u, \cdot) \ du \ = \ \{ \tau_-^{+ \frac{1}{2}} \} 
\ee
and 
\be \label{Sigma1234beta}
\int_u \mathcal{S}_{\beta} (u, \cdot)  \ du \ = \  \{  \tau_-^{\beta} \} 
\ee
Then the following is a consequence of (\ref{Q3Sigma1234*}) and proposition \ref{Rholimits1}: 
\[
\mathcal{Q} \ = \ \int_u \mathcal{S}_{\frac{1}{2}} (u, \cdot) \ du \ + \ \int_u \mathcal{S}_{\beta} (u, \cdot)  \ du \ + \ l.o.t. 
\]
Therefore, the first term on the right hand side of (\ref{limitstructuresdawn}) is given by (\ref{Sigma1234}) and the second term on the right hand side of (\ref{limitstructuresdawn}) is given by (\ref{Sigma1234beta}). 
We have shown how these quantities are rooted in $\sigma_3$. 
On the right hand side of (\ref{limitstructuresdawn}) 
denote 
$(\{ \tau_-^{+ \frac{1}{2}} \} \ + \ \{  \tau_-^{\beta} \})$ by $ \mathcal{Q}_{\sigma_1}$ to express this connection with $\sigma_3$. 
In other words, the terms of order $O( \tau_-^{\alpha})$ with $0 < \alpha \leq \frac{1}{2}$ originate from the integral of the limits of the corresponding $\sigma_3$ parts. 

Next, we investigate the third term on the right hand side of (\ref{limitstructuresdawn}). For this purpose, recall proposition \ref{rho_3etal****}. Take the limits in (\ref{defQ3****2}) and (\ref{defQ3*2}). 
We conclude that $\{ \mathcal{F} (u, \cdot) \}$ depends on $u$ but does not decay in $u$. More precisely, the integral in $u$ of the corresponding limits 
$\lim_{C_u, t \to \infty} (r^3 \sigma_f)$, respectively $\lim_{C_u, t \to \infty} (r^3 g)$ 
of the terms 
$\sigma_f$, respectively $g$ in (\ref{slandi2}) make up $\mathcal{F} (u, \cdot)$. We denote by $ \mathcal{Q}_{\sigma_2 }$ the parts of $\mathcal{F} (u, \cdot)$ that are in this way rooted in $\sigma_3$ and by $- \frac{1}{2} G$ the parts rooted in $\frac{\partial}{\partial u} (\hat{\chi} \wedge \hat{\underline{\chi}})$. 

Likewise, for the fourth term on the right hand side of (\ref{limitstructuresdawn}), 
recall proposition \ref{rho_3etal****} and 
take the corresponding limits in (\ref{defQ3****2}) and (\ref{defQ3*2}). It follows that $\{  \tau_-^{-  \gamma} \}$ is sourced by terms that originate in $\sigma_3$ as well as by terms that originate in $\frac{\partial}{\partial u} (\hat{\chi} \wedge \hat{\underline{\chi}})$. As $\{  \tau_-^{-  \gamma} \}$ decay in $u$, they will not contribute to the effects to be studied. 

Finally, $C_{\mathcal{Q}}$ in  (\ref{limitstructuresdawn}) is the said constant that will cancel later. 

This proves proposition \ref{Qres}.

Utilizing the new structures of proposition \ref{Qres} and proof, we write 
\be \label{limitstructures2dawn}
\mathcal{Q} (u, \cdot) \ = \ \mathcal{Q}_{\sigma_1}  (u, \cdot) \ + \ \mathcal{Q}_{\sigma_2 }  (u, \cdot) \ - \frac{1}{2}  \ G  (u, \cdot) \ + \ C_{\mathcal{Q}}  \ + \ l.o.t. 
\ee
We emphasize that $\mathcal{Q}_{\sigma_2 }$ and $G$ depend on $u$. 
This is very different from AF systems of the order $O(r^{-1})$, 
where 
it is always 
$\hat{\chi} \wedge \hat{\underline{\chi}} = O(r^{-3} \tau_-^{- \alpha})$ with $\alpha > 0$. In the latter situations, the quantities $\hat{\chi}$ as well as $\hat{\underline{\chi}}$ take limits at $\mathcal{I}^+$ such that 
the wedge product of their limits vanishes as $|u| \to \infty$. 
Therefore $G^-=G^+=0$. \\ 

{\bf Remark:} From the above it follows that $\sigma$ may have terms of the order $O(r^{-3})$. The latter is impossible for AF systems of stronger decay such as (CKvac) or (Mvac) spacetimes. In those stronger situations one has $\sigma = O(r^{-3} \tau_-^{- \alpha})$ with $\alpha > 0$. However, note also that in those systems it emerges naturally that $\rho = O(r^{-3})$. \\

Next, consider ({\ref{Q3limit1234}) and employ again the relations between $\hat{\chi}$, $\hat{\underline{\chi}}$ and $\underline{\beta}$ as well as the corresponding limits 
(\ref{Ldchibund1}) and (\ref{Lchihat}) 
to find 
\be \label{1supergoldsigma}
\mathcal{Q}_3 \ = \ - 2 \ \clap \ \ \dlap Chi_3
\ee
Integrating (\ref{1supergoldsigma}) with respect to $u$ yields 
\be \label{2supergoldsigmausw}
(\mathcal{Q}^- - \mathcal{Q}^+) 
  \ = \ 
\clap \ \  \dlap (Chi^- - Chi^+) 
\ee
Using (\ref{limitstructures2dawn}) we have 
\be \label{2supergoldsigma}
(\mathcal{Q}_{\sigma_1}^- - \mathcal{Q}_{\sigma_1}^+) 
 \ + \ (\mathcal{Q}_{\sigma_2}^- - \mathcal{Q}_{\sigma_2}^+) 
\ - \  \frac{1}{2} (G^- - G^+) 
  \ = \ 
\clap \ \  \dlap (Chi^- - Chi^+) 
\ee
We have derived new structures. 
We observe the following important fact for $(\mathcal{Q}^- - \mathcal{Q}^+)$, respectively $\clap \ \ \dlap (Chi^- - Chi^+)$ for (Bvac) as well as (Avac) spacetimes: Recall the concept of limits at null infinity $\mathcal{I^+}$ introduced in section \ref{limits12}. 
Fix a point on the sphere $S^2$ at fixed $u_0$. Then consider $\mathcal{Q}(u_0)$. Next, take $\mathcal{Q}(u)$ at the corresponding point for some value of $u \neq u_0$. Keep $u_0$ fixed and let $u$ tend to $+ \infty$, respectively to $- \infty$. Then the difference 
$\mathcal{Q}(u) - \mathcal{Q}(u_0)$ is not finite, but it grows with $|u|^{+ \frac{1}{2}}$ for (Bvac) as well as (Avac) spacetimes. 
A corresponding argument holds for $Chi(u) - Chi(u_0)$. 
Recall from (\ref{limitstructuresdawn}) that we collected the diverging terms in $\mathcal{Q}_{\sigma_1}$. 
In (\ref{2supergoldsigma}), we see that 
$(\mathcal{Q}_{\sigma_2}^- - \mathcal{Q}_{\sigma_2}^+)$ is finite and 
$(G^- - G^+)$ is finite. 

Besides emphasizing all these new structures, special importance belongs to the new result that for fixed $u_0$ the difference $\mathcal{Q} (u) - \mathcal{Q} (u_0)$ grows like $|u|^{\frac{1}{2}}$ as $|u| \to \infty$.

At this point, recall the system (\ref{Phi1}) - (\ref{Phi2}). In this section, we are considering the general case where $\nlap^{\perp} \Psi \neq 0$. 
In particular, 
instead of (\ref{Phi1}) we have (\ref{Psi1}) together with the new equation (\ref{Psi2}) with non-trivial right hand side.   
More precisely, there exist functions 
$\Phi$ and $\Psi$ such that 
$\dlap (Chi^- - Chi^+) = \nlap \Phi + \nlap^{\perp} \Psi$. 
Let $Z := \dlap (Chi^- - Chi^+)$. 
Note that then the following holds: 
\[
\dlap Z \ = \ \slap \Phi \ \ \ \ , \ \ \ \ \ \clap \ \ Z \ = \ \slap \Psi \ \ . 
\]
We obtain the new system 
\bea
 \dlap (Chi^- - Chi^+) & = & \nlap \Phi + \nlap^{\perp} \Psi  \label{Psi1} \\ 
\clap \ \  \dlap (Chi^- - Chi^+) & = & \slap \Psi  \nonumber \\ 
& = & (\mathcal{Q} - \bar{\mathcal{Q}})^- - (\mathcal{Q} - \bar{\mathcal{Q}})^+   \label{Psi2} \\ 
\dlap \dlap (Chi^- - Chi^+) & = & \slap \Phi  \nonumber \\ 
& = & (\mathcal{P} - \bar{\mathcal{P}})^- - (\mathcal{P} - \bar{\mathcal{P}})^+ \nonumber \\ 
& & - 2 (F - \bar{F})  \label{Phi33} 
\eea
That is, we have the following system on $S^2$, that is solved by Hodge theory. 
\bea
 \dlap (Chi^- - Chi^+) & = & \nlap \Phi + \nlap^{\perp} \Psi  \label{77Psi1**} \\ 
\clap \ \  \dlap (Chi^- - Chi^+) & = & \slap \Psi  \nonumber \\ 
& = & (\mathcal{Q}_{\sigma_1} - \bar{\mathcal{Q}}_{\sigma_1})^- - (\mathcal{Q}_{\sigma_1} - \bar{\mathcal{Q}}_{\sigma_1})^+ \nonumber \\ 
& & + (\mathcal{Q}_{\sigma_2} - \bar{\mathcal{Q}}_{\sigma_2})^- - (\mathcal{Q}_{\sigma_2} - \bar{\mathcal{Q}}_{\sigma_2})^+ \nonumber \\ 
& & - \frac{1}{2} (G - \bar{G})^- +  \frac{1}{2} (G - \bar{G})^+   \label{77Psi2**} \\ 
\dlap \dlap (Chi^- - Chi^+) & = & \slap \Phi  \nonumber \\ 
& = & (\mathcal{P}_{\rho_1 } - \bar{\mathcal{P}}_{\rho_1 })^- - (\mathcal{P}_{\rho_1 } - \bar{\mathcal{P}}_{\rho_1 })^+ \nonumber \\ 
& & (\mathcal{P}_{\rho_2 } - \bar{\mathcal{P}}_{\rho_2 })^- - (\mathcal{P}_{\rho_2 } - \bar{\mathcal{P}}_{\rho_2 })^+ \nonumber \\ 
& & - 2 (F - \bar{F})  \label{77Phi2} \\ 
& & - \frac{1}{2} (D - \bar{D})^- +  \frac{1}{2} (D - \bar{D})^+ \nonumber
\eea

For (Bvac) spacetimes the following holds, and is further conjectured for (Avac) spacetimes. 
We conclude that 
there is the {\bf new magnetic memory effect} growing with $|u|^{\frac{1}{2}}$ sourced by $\mathcal{Q}$ and finite contributions from both $\mathcal{Q}$ and $G$. Moreover, $\mathcal{Q}$ has further diverging terms at lower order. 
In addition, we have the {\bf electric memory}, previously established. This electric part is growing with $|u|^{\frac{1}{2}}$ sourced by 
$\mathcal{P}$, and finite contributions from $\mathcal{P}$, $D$ and from $F$ (the latter may be unbounded for (Avac)). 

The right hand side of (\ref{Psi2}), respectively (\ref{77Psi2**}), being non-zero, is a direct consequence from the more general data considered here. Thus, the magnetic memory occurs naturally in (Bvac) spacetimes and is conjectured for (Avac) spacetimes. 
The fact of $\clap \ \ \dlap (Chi^- - Chi^+)$ being non-trivial allows for the magnetic structures to appear in gravitational radiation and to enter the permanent changes of the spacetime.

It follows that 
gravitational radiation in AF spacetimes of types (Bvac) generates the growing memories derived in sections \ref{FNI} and \ref{new}. The corresponding statement is conjectured for (Avac) spacetimes.

We have {\bf proven} the following theorem. 

\begin{The} \label{theelmagn***}
The following holds for (Bvac) spacetimes. 

$(Chi^- - Chi^+)$ is determined by equation (\ref{77Psi1**}) on $S^2$ where $\Psi$ solves (\ref{77Psi2**}) and $\Phi$ solves (\ref{77Phi2}). 

\end{The}

\begin{conj} \label{conjtheelmagn***}
The statement of theorem \ref{theelmagn***} holds correspondingly for (Avac) spacetimes. 
\end{conj}

It is important to emphasize that for (CKvac) or (Mvac) spacetimes magnetic memory does {\bf not} occur. 

The present situation of spacetimes with slow decay provides a more general framework in which $\sigma$ as well as $(\hat{\chi} \wedge \hat{\underline{\chi}})$ feature terms that are not present in (CKvac) nor (Mvac) spacetimes. 

The magnetic memory together with the new structures emerge naturally for the more general spacetimes.

\subsection{Past Null Infinity and Memories}
\label{PNI}

Similarly to \cite{lydia1}, \cite{lydia2} for the future development of the initial data therein, we can perform the extension towards the past to obtain the following decay behavior at infinity: 
\bea
\alpha \ & = & \ O \ ( r^{- 1} \ \tau_+^{- \frac{3}{2}})   \label{resa2} \\ 
\beta \ & = & \ O \ ( r^{- 2} \ \tau_+^{- \frac{1}{2}})  \label{resb2}   \\ 
\rho , \ \sigma , \ \underline{\alpha} , \ \underline{\beta}  \ & = & \ o \ (r^{- \frac{5}{2}})   \label{resrsuaub2}   \\ 
\hat{\chi}  \ & = & \ O \ (r^{-1} \tau_+^{- \frac{1}{2}})  \label{reshatchi2}  \\ 
\underline{\hat{\chi}}  \ & = & \ o \ (r^{- \frac{3}{2}})   \label{resuhatchi2}   \\ 
tr \chi - \overline{tr \chi} \ & = & \ O \ (r^{- 2})   \label{restrchi2}  \\ 
\zeta  \ & = & \ o \ (r^{- \frac{3}{2}})  \label{resz2}   \\ 
\underline{\zeta}  \ & = & \ o \ (r^{- \frac{3}{2}})  \label{resuz2}   \\ 
K  - \frac{1}{r^2}  \ & = & \  o \ (r^{- \frac{5}{2}})   \label{resK2} 
\eea
Further, we have 
\beas
\theta \ & = & \ 
O(r^{-1} \tau_+^{- \frac{1}{2}}) \\ 
\hat{\eta}  \ & = & \ 
O(r^{-1} \tau_+^{- \frac{1}{2}}) \\ 
\epsilon \ & = & \ o(r^{- \frac{3}{2}})  \\ 
\delta \ & = & \ o(r^{- \frac{3}{2}})  \\ 
\eeas

Equation (\ref{schihat1}) yields
\be \label{uuuchihat2}
\frac{\partial}{\partial \underline{u}} \hat{\chi} \ = \ 
\frac{1}{2} \alpha \ + \ l.o.t. 
\ee
and equation (\ref{suchihat1}) yields
\be \label{uuuuchihatbar2}
\frac{\partial}{\partial \underline{u}} \underline{\hat{\chi}} \ = \ 
- \frac{1}{4} tr \underline{\chi} \hat{\chi}  \ + \ l.o.t. 
\ee

We have the following limits at $\mathcal{I}^-$. 
\bea
- \frac{1}{2} \lim_{C_{\underline{u}}, t \to - \infty } r \hat{\chi} \  =  \ \lim_{C_{\underline{u}}, t \to - \infty } r \hat{\eta}  \ & = & \ \Sigma(\underline{u})
\eea

The normalized curvature components $r \alpha  $, $r^{2} \beta$ have limits on $C_{\underline{u}}$ as $t\rightarrow - \infty $: 
\begin{eqnarray*}
\lim_{C_{\underline{u}},t\rightarrow - \infty }r {\alpha } 
&=&A \left( \underline{u},\cdot \right) ,\, \ \ \ \ \ \ \ \ \ \ \ \
\lim_{C_{\underline{u}},t\rightarrow - \infty }\,r^{2} {\beta } 
= \underline{B}\left( \underline{u},\cdot \right) \ , 
\end{eqnarray*}
where the limits are on $S^{2}$ and depend on $\underline{u}$. These limits satisfy 
\begin{eqnarray*}
\left| A \left( \underline{u},\cdot \right) \right| &\leq &C\left( 1+\left| \underline{u} \right|
\right) ^{-3/2}\, \, \ \ \ \ \ \ \ \ \ \ \ \ \left| {B}\left( \underline{u},\cdot
\right) \right| \leq C\left( 1+\left| \underline{u} \right| \right) ^{-1/2}  . \ \
\end{eqnarray*}

Then, we define 
\bea
Xi_4 & : =  & \lim_{C_u, t \to \infty} \big{(}  r^2 \frac{\partial}{\partial \underline{u}}  \underline{\hat{\chi}} \label{defXi4*1} \big{)} \\ 
Xi & : =  & \int_{ \underline{u}} Xi_4 \ d  \underline{u}  \label{defXi4*2}
\eea

Note that $Xi$ is defined on $S^2 \times \real$ up to an additive function on $S^2$ independent of $\underline{u}$. 
Later, when taking the integral $\int_{- \infty}^{+ \infty} Xi_4 \ d  \underline{u}$ that additive function will cancel. 

From (\ref{dchib1}) it follows that 
\be \label{LBSF1}
B \ = \ 2 \dlap \Sigma 
\ee

Equation (\ref{uuuchihat2}) yields 
\be \label{LSigma22}
\frac{\partial}{\partial \underline{u}} \Sigma \ = \ - \frac{1}{4} A 
\ee

and equation (\ref{uuuuchihatbar2}) gives 
\be \label{LSigma33}
Xi_4  \ = \ - \Sigma 
\ee

Consider the Bianchi equation 
\bea 
\Dlap_4  \rho \ + \ \frac{3}{2} tr \chi \rho \ & = & \ 
\dlap \beta -  \frac{1}{2} \underline{\hat{\chi}} \alpha \ + \ 
\epsilon \beta + 2 \underline{\zeta} \beta 
  \label{Bianchiturho4}  
 \eea
Denote $\rho_4 : = \Dlap_4  \rho \ + \ \frac{3}{2} tr \chi \rho $ and write 
\be \label{Bianchiturho5}
\rho_4  \  =  \ 
\underbrace{\dlap \beta}_{ O(r^{-3} \tau_+^{- \frac{1}{2}})} -  \underbrace{\frac{1}{2} \underline{\hat{\chi}} \cdot \alpha}_{= O(r^{- \frac{5}{2}} \tau_+^{- \frac{3}{2}})} \ + \ l.o.t. 
\ee
A short computation gives 
\[
\rho_4  \  =  \ 
\underbrace{\dlap \beta}_{= O(r^{-3} \tau_+^{- \frac{1}{2}})} -
 \underbrace{\frac{\partial}{\partial \underline{u}} (\hat{\chi} \cdot \hat{\underline{\chi}})}_{ = O(r^{- \frac{5}{2}} \tau_+^{- \frac{3}{2}})} 
 - \underbrace{ \frac{1}{4} tr \underline{\chi} | \hat{\chi} |^2 }_{= O(r^{-3} \tau_+^{-1})}
\]
Thus 
\be \label{rho4*}
\rho_4  + \frac{\partial}{\partial \underline{u}} (\hat{\chi} \cdot \hat{\underline{\chi}}) \ = \ 
\dlap \beta - 
 \frac{1}{4} tr \underline{\chi} | \hat{\chi} |^2 
  \ =  \ O(r^{-3} \tau_+^{- \frac{1}{2}})
\ee

We introduce the following notation for the corresponding limit of the left hand side of (\ref{rho4*}): 
\bea
\mathcal{P}_{4} & : =  & \lim_{C_{\underline{u}}, t \to - \infty} r^3 \big{(} \rho_4  + \frac{\partial}{\partial \underline{u}} (\hat{\chi} \cdot \hat{\underline{\chi}}) \label{defrho3*limit**188} \\ 
\mathcal{P} & : =  & \int_{\underline{u}} \mathcal{P}_4 \ d \underline{u} \label{defP3*2limit**188} 
\eea

Investigating the structures of $\mathcal{P}$ yields for $0 < \beta < \frac{1}{2}$ and $\gamma > 0$, 
\bea 
\mathcal{P} \ & = & \ \underbrace{\{ \tau_+^{+ \frac{1}{2}} \} \ + \ \{  \tau_+^{\beta} \}}_{ = \mathcal{P}_{\rho_1}}  \ + \ 
\underbrace{\{ \mathcal{F} (\underline{u}, \cdot) \}}_{= \mathcal{P}_{\rho_2 } - \frac{1}{2} D} \ + \ \{  \tau_+^{-  \gamma} \}   \ + \ 
C_{\mathcal{P}} 
\label{limitstructures44}
\eea
where $\mathcal{F} (\underline{u}, \cdot) \leq C$. 
And $C_{\mathcal{P}}$ is an 
additive function on $S^2$ not depending on $\underline{u}$. Later, when taking the integral 
$\int_{- \infty}^{+ \infty} \mathcal{P}_4 \ d \underline{u}$, the term $C_{\mathcal{P}}$ will cancel. 
It follows that the terms of order $O( \tau_+^{\alpha})$ with $0 < \alpha \leq \frac{1}{2}$ originate from the integral of the limits of the $\rho_4$ part. We denote this part by $\mathcal{P}_{\rho_1}$. In (\ref{limitstructures44}), $\mathcal{F} (\underline{u}, \cdot)$ 
comprises the pieces of the integral sourced by the limits of the terms of order $O(r^{-3})$ in (\ref{rho4*}). 
It has components sourced by $\rho_4$ as well as by $\frac{\partial}{\partial \underline{u}} (\hat{\chi} \cdot \hat{\underline{\chi}})$, we denote the former by $\mathcal{P}_{\rho_2 }$ and the latter by $D$. 
Thus we have 
\be \label{limitstructures244}
\mathcal{P} \ = \ \mathcal{P}_{\rho_1} \ + \ \mathcal{P}_{\rho_2 } \ - \frac{1}{2}  \ D  \ + \ 
C_{\mathcal{P}} 
 \ + \ l.o.t. 
\ee
It also follows that $\mathcal{P}_{\rho_2 }$ as well as $D$ are depending on $\underline{u}$.

Multiply equation (\ref{rho4*}) by $r^3$ and take the limit on $C_{\underline{u}}$ as $t \to - \infty$ to obtain 
\be \label{LCrho4**}
\mathcal{P}_4 \ = \ 
 - 2 \dlap \dlap Xi_4 + 2 | \Sigma |^2 
\ee

We obtain 
\be \label{supergold2-+} 
\dlap \dlap (Xi^+ - Xi^-) \ = \ 
- (\mathcal{P}^+ - \mathcal{P}^- )
+ \int_{- \infty}^{+ \infty} | \Sigma |^2 d \underline{u} 
\ee
That is, there exists a function $\Phi$ on $S^2$ such that 
\bea
 \dlap (Xi^+ - Xi^-)  \ & = & \  \nlap \Phi + X    \label{supergoldpreambel1} \\ 
\dlap \dlap (Xi^+ - Xi^-) \ & = & \ \slap \Phi  \nonumber \\ 
\ & = & \ 
- (\mathcal{P}_{\rho_1}^+ - \mathcal{P}_{\rho_1}^- ) 
- (\mathcal{P}_{\rho_2}^+ - \mathcal{P}_{\rho_2}^- )
+ \int_{- \infty}^{+ \infty} | \Sigma |^2 d \underline{u} 
- \frac{1}{2} ( D^+ - D^- )  \nonumber \\ 
\label{supergold2} 
\eea
with $X = \nlap^{\perp} \Psi$ for a function $\Psi$ whose Laplacian $\slap \Psi = \clap \ \  \dlap (Xi^+ - Xi^-)$. 
However, similarly as in section \ref{FNI}, we want to first explain the new structures for electric memory more clearly by putting the contribution from $\Psi$ to zero for the moment. Then we treat the most general case where $\nlap^{\perp} \Psi \neq 0$ to obtain the complete general result and the full set of equations (\ref{PPsi1})-(\ref{PPhi33}).

For (Bvac) spacetimes, in (\ref{supergold2}) the integral $\int_{- \infty}^{+ \infty} | \Sigma |^2 d \underline{u}$ is finite, in fact it is borderline in view of how $\Sigma$ decays in $\underline{u}$. The data from \cite{lydia1, lydia2} guarantees this term to be finite, that is borderline. Relaxing this data will compel this integral to diverge. The latter happens for (Avac) spacetimes. Other than for the null memory, the behavior of the remaining terms in (\ref{supergold2}) holds for both (Bvac) as well as (Avac) spacetimes. 

We see that incoming radiation causes an electric memory effect growing like $\sqrt{ | \underline{u} | }$ due to $\mathcal{P}_{\rho_1}$ (ordinary memory). The memory has a finite contribution from $ \int_{- \infty}^{+ \infty} | \Sigma |^2 d \underline{u}$ (null memory). Moreover, it has a finite contribution from the second term 
as well as the last term on the right hand side of (\ref{supergold2}).
As long as radiation comes in, this memory keeps growing at the rate $\sqrt{ | \underline{u} | }$. This comprises the electric gravitational memory effect at 
$\mathcal{I}^-$. 

We have {\bf proven} the following theorem. 

\begin{The} \label{incomingvogel}
The following holds for (Bvac) spacetimes. 

If $\nlap^{\perp} \Psi \equiv 0$, then  
$(Xi^+ - Xi^-)$ is determined by equation (\ref{supergoldpreambel1}) on $S^2$ where $\Phi$ is the solution with vanishing mean of (\ref{supergold2}). 

\end{The}

\begin{conj} 
The statement of theorem \ref{incomingvogel} holds correspondingly for (Avac) spacetimes.

\end{conj}

If we ``turn off" the incoming radiation, then in particular $\Sigma \equiv 0$, and each limit in equation (\ref{supergold2}) is identically zero. Therefore, under the no incoming radiation condition, there is no memory at $\mathcal{I}^-$. 

Comparing this to future null infinity $\mathcal{I}^+$, we find that the gravitational wave memory effect at $\mathcal{I}^+$ is always there, as long as we have outgoing radiation; whereas only for incoming gravitational waves we find a counterpart at $\mathcal{I}^-$. The latter disappears for systems without incoming radiation. 
Thus, the same initial data always creates an outgoing memory at $\mathcal{I}^+$. 

Let us also investigate the magnetic counterpart at $\mathcal{I}^-$. Now, we treat the most general case with $\nlap^{\perp} \Psi \neq 0$. 

Denote $\sigma_4 = \Dlap_4 \sigma + \frac{3}{2} tr \chi \sigma$. 
Next, we consider the Bianchi equation 
\[
\sigma_4 \ = \ - \clap \ \ \beta + \frac{1}{2} \underline{\hat{\chi}} \ ^* \alpha - \epsilon \ ^* \beta - 2 \underline{\zeta} \ ^* \beta 
\]
The lower order terms decay to zero as we approach null infinity, thus we focus on the relevant part of the equation: 
\be \label{s414}
\sigma_4 \ = \ - \clap \ \ \beta + \frac{1}{2} \underline{\hat{\chi}} \ ^* \alpha + l.o.t. 
\ee
and compute 
\be \label{news414}
\sigma_4 - \frac{\partial }{\partial \underline{u}} \big{(} \underline{\hat{\chi}} \wedge \hat{\chi} \big{)} \ = \ 
- \clap \ \ \beta \ = \ O(r^{-3} \tau_+^{- \frac{1}{2}})
\ee
Next, introduce the following notation for the corresponding limit of the left hand side of (\ref{news414}): 
\bea
\mathcal{Q}_{4} & : =  & \lim_{C_{\underline{u}}, t \to - \infty} r^3 \big{(} \sigma_4  - \frac{\partial}{\partial \underline{u}} (\hat{\chi} \wedge \hat{\underline{\chi}}) \big{)} \label{news44*1} \\ 
\mathcal{Q} & : =  & \int_{\underline{u}} \mathcal{Q}_4 \ d \underline{u} \label{news442**1} 
\eea

Investigating the structures of $\mathcal{Q}$ yields for $0 < \beta < \frac{1}{2}$ and $\gamma > 0$, 
\bea 
\mathcal{Q} \ & = & \ \underbrace{\{ \tau_+^{+ \frac{1}{2}} \} \ + \ \{  \tau_+^{\beta} \}}_{ = \mathcal{Q}_{\rho_1}}  \ + \ 
\underbrace{\{ \mathcal{F} (\underline{u}, \cdot) \}}_{= \mathcal{Q}_{\rho_2 } - \frac{1}{2} G} \ + \ \{  \tau_+^{-  \gamma} \} \ + \ C_{\mathcal{Q}} 
\label{limitstructures4455}
\eea
where $\mathcal{F} (\underline{u}, \cdot) \leq C$. And 
$C_{\mathcal{Q}}$ is an 
additive function on $S^2$ not depending on $\underline{u}$. Later, when taking the integral 
$\int_{- \infty}^{+ \infty} \mathcal{Q}_4 \ d \underline{u}$, the term $C_{\mathcal{Q}}$ will cancel. 
It follows that the terms of order $O( \tau_+^{\alpha})$ with $0 < \alpha \leq \frac{1}{2}$ originate from the integral of the limits of the $\sigma_4$ part. We denote this part by $\mathcal{Q}_{\rho_1}$. In (\ref{limitstructures4455}), $\mathcal{F} (\underline{u}, \cdot)$ 
comprises the pieces of the integral originating from the limits of the terms of order $O(r^{-3})$ in (\ref{news414}). 
It has components sourced by $\sigma_4$ as well as by $\frac{\partial}{\partial \underline{u}} (\hat{\chi} \wedge\hat{\underline{\chi}})$, we denote the former by $\mathcal{Q}_{\rho_2 }$ and the latter by $G$. 
Thus we have 
\be \label{limitstructures24455}
\mathcal{Q} \ = \ \mathcal{Q}_{\rho_1} \ + \ \mathcal{Q}_{\rho_2 } \ - \frac{1}{2}  \ G \ + \ C_{\mathcal{Q}} 
 \ + \ l.o.t. 
\ee
It also follows that $\mathcal{Q}_{\rho_2 }$ as well as $G$ are depending on $\underline{u}$.

Multiply equation (\ref{news414}) by $r^3$ and take the limit on $C_{\underline{u}}$ as $t \to - \infty$ to obtain 
\be \label{LCsigma4**}
\mathcal{Q}_4 
\ = \ 
-   \clap \ \ B 
\ee
That is 
\be
\mathcal{Q}_4 
\ = \ 2 \ \clap \ \ \dlap Xi_4 
\ee
We compute 
\be \label{**supergoldsigma**255}
\clap \ \ \dlap ( Xi^+ - Xi^- ) \ = \ 
(\mathcal{Q}^+ - \mathcal{Q}^-) 
\ee
That is 
\be \label{**supergoldsigma**2}
\clap \ \ \dlap ( Xi^+ - Xi^- ) \ = \ 
(\mathcal{Q}_{\rho_1}^+ - \mathcal{Q}_{\rho_1}^-)  
- (\mathcal{Q}_{\rho_2}^+ - \mathcal{Q}_{\rho_2}^-)
- \frac{1}{2} (G^+ - G^-)  
\ee
These results hold for (Bvac) spacetimes and are conjectured for (Avac) spacetimes.

Similar to the situation at future null infinity $\mathcal{I}^+$ we have a Hodge system at past null infinity $\mathcal{I}^-$. 
Let $Z := \dlap (Xi^+ - Xi^-)$. 
Then the following holds: 
\[
\dlap Z \ = \ \slap \Phi \ \ \ \ , \ \ \ \ \ \clap \ \ Z \ = \ \slap \Psi \ \ . 
\]
Again, 
Hodge theory provides the solution to the following system. 
\bea
 \dlap (Xi^+ - Xi^-) & = & \nlap \Phi + \nlap^{\perp} \Psi  \label{PPsi1} \\ 
\clap \ \  \dlap (Xi^+ - Xi^-) & = & \slap \Psi  \nonumber \\ 
& = & (\mathcal{Q} - \bar{\mathcal{Q}})^+ - (\mathcal{Q} - \bar{\mathcal{Q}})^-   \label{PnewPsi2} \\ 
\dlap \dlap (Xi^+ - Xi^-) & = & \slap \Phi  \nonumber \\ 
& = & - (\mathcal{P} - \bar{\mathcal{P}})^+ + (\mathcal{P} - \bar{\mathcal{P}})^- \nonumber \\ 
& & + 2 (F_P - \bar{F_P})  \label{PPhi33} \\ \nonumber
\eea
where 
$F_P$ is defined to be 
\[
F_P (\cdot) = \frac{1}{2} \int_{- \infty}^{+ \infty} | \Sigma |^2 d \underline{u}  
\] 
That is 
\bea
 \dlap (Xi^+ - Xi^-) & = & \nlap \Phi + \nlap^{\perp} \Psi  \label{55PPsi1} \\ 
\clap \ \  \dlap (Xi^+ - Xi^-) & = & \slap \Psi  \nonumber \\ 
& = & (\mathcal{Q}_{\sigma_1} - \bar{\mathcal{Q}}_{\sigma_1})^+ - (\mathcal{Q}_{\sigma_1} - \bar{\mathcal{Q}}_{\sigma_1})^- \nonumber \\ 
& & + (\mathcal{Q}_{\sigma_2} - \bar{\mathcal{Q}}_{\sigma_2})^+ - (\mathcal{Q}_{\sigma_2} - \bar{\mathcal{Q}}_{\sigma_2})^- \nonumber \\ 
& & - \frac{1}{2} (G - \bar{G})^+ +  \frac{1}{2} (G - \bar{G})^-   \label{55PnewPsi2} \\ 
\dlap \dlap (Xi^+ - Xi^-) & = & \slap \Phi  \nonumber \\ 
& = & - (\mathcal{P}_{\sigma_1} - \bar{\mathcal{P}}_{\sigma_1})^+ + (\mathcal{P}_{\sigma_1} - \bar{\mathcal{P}}_{\sigma_1})^- \nonumber \\ 
& &  - (\mathcal{P}_{\sigma_2} - \bar{\mathcal{P}}_{\sigma_2})^+ + (\mathcal{P}_{\sigma_2} - \bar{\mathcal{P}}_{\sigma_2})^- \nonumber \\ 
& & + 2 (F_P - \bar{F_P})  \label{55PPhi33} \\ 
& & - \frac{1}{2} (D - \bar{D})^+ +  \frac{1}{2} (D - \bar{D})^- \nonumber 
\eea
We conclude that there is {\bf electric memory} growing with $|\underline{u}|^{\frac{1}{2}}$ sourced by $\mathcal{P}$ and finite contribution from $F_P$. (Again, the latter may not be bounded for (Avac) spacetimes.) And there is also {\bf magnetic memory} growing with $|\underline{u}|^{\frac{1}{2}}$ sourced by $\mathcal{Q}$. 

We have {\bf proven} the following theorem. 

\begin{The} \label{incomingbussard}
The following holds for (Bvac) spacetimes. 

$(Xi^+ - Xi^-)$ is determined by equation (\ref{55PPsi1}) on $S^2$ where $\Psi$ solves (\ref{55PnewPsi2}) and $\Phi$ solves (\ref{55PPhi33}). 

\end{The}

\begin{conj} 
The statement of theorem \ref{incomingbussard} holds correspondingly for (Avac) spacetimes. 

\end{conj}

Incoming radiation can include gravitational waves from other non-cosmological sources or they can be of primordial nature originating in the early universe.

Under the NIRC all memories at $\mathcal{I}^-$ vanish.

\subsection{Gravitational Radiation and Memory}
\label{beides}

We collect the gravitational wave memory effects derived above for (Bvac) spacetimes and conjectured for (Avac) spacetimes. 

At future null infinity $\mathcal{I}^+$ the electric and magnetic memories are determined by the following system, where the notation is as introduced in sections \ref{FNI} and \ref{new}: 
\beas
 \dlap (Chi^- - Chi^+) & = & \nlap \Phi + \nlap^{\perp} \Psi  \\ 
\clap \ \  \dlap (Chi^- - Chi^+) & = & \slap \Psi  \nonumber \\ 
& = & (\mathcal{Q} - \bar{\mathcal{Q}})^- - (\mathcal{Q} - \bar{\mathcal{Q}})^+    \\ 
\dlap \dlap (Chi^- - Chi^+) & = & \slap \Phi  \nonumber \\ 
& = & (\mathcal{P} - \bar{\mathcal{P}})^- - (\mathcal{P} - \bar{\mathcal{P}})^+ \nonumber \\ 
& & - 2 (F - \bar{F}) 
\eeas
At past null infinity $\mathcal{I}^-$ the electric and magnetic memories are determined by the following system, where the notation is as introduced in section \ref{PNI}: 
\beas
 \dlap (Xi^+ - Xi^-) & = & \nlap \Phi + \nlap^{\perp} \Psi   \\ 
\clap \ \  \dlap (Xi^+ - Xi^-) & = & \slap \Psi  \nonumber \\ 
& = & (\mathcal{Q} - \bar{\mathcal{Q}})^+ - (\mathcal{Q} - \bar{\mathcal{Q}})^-   \\ 
\dlap \dlap (Xi^+ - Xi^-) & = & \slap \Phi  \nonumber \\ 
& = & - (\mathcal{P} - \bar{\mathcal{P}})^+ + (\mathcal{P} - \bar{\mathcal{P}})^- \nonumber \\ 
& & + 2 (F_P - \bar{F_P})   \\ \nonumber
\eeas

We recall the important structures (\ref{limitstructures2}), (\ref{limitstructures2dawn}), (\ref{limitstructures244}), (\ref{limitstructures24455}). 

The memory at future null infinity $\mathcal{I}^+$ is caused by outgoing radiation sourced by $\Xi$, $\mathcal{P}$, and $\mathcal{Q}$. The memory at past null infinity $\mathcal{I}^-$ is caused by incoming radiation sourced by $\Sigma$, $\mathcal{P}$, and $\mathcal{Q}$. 

Under the NIRC all memories at $\mathcal{I}^-$ vanish. The memories at $\mathcal{I}^+$ always occur. \\

\section{Einstein-Null-Fluid Equations for Neutrinos} 
\label{neutrinos}

Non-isotropic distribution of neutrinos source the radiation. 
In this section, we investigate spacetimes of the types (B-Tneutrinos) and (A-Tneutrinos). 
We shall prove new results for (B-Tneutrinos) spacetimes and provide evidence for new conjectures in (A-Tneutrinos) spacetimes. 
They describe neutrino distributions with slow decay towards infinity. 
In particular, they are not stationary outside a compact set. The neutrinos are modeled by a null fluid coupled to the Einstein equations (\ref{ENF}). Based on these results, we will compute new structures in the gravitational radiation from these sources and derive the new effects for these spacetimes.  \\

We recall the Einstein-null-fluid equations describing neutrinos (\ref{ENF}) 
\[
R_{\mu \nu}  \ = \ 8 \pi \ T_{\mu \nu} \ . 
\]

\subsection{Results and Setting}
\label{neutrinosintro}

Our new results, described in the previous sections, emerge from the Einstein vacuum equations (\ref{EV}). 
In this section, we couple the Einstein equations to a source. 

A possible source for this radiation and the new memory effects are neutrinos moving non-isotropically throughout large regions. In particular, such a source is not stationary outside a compact set, but its distribution may linger with a slow decay over huge regions for a long time. 

Neutrinos move at almost the speed of light and have a very small mass. 
Together with A. Tolish, D. Garfinkle, and R. Wald we showed in \cite{lbatdgbw} that massive particles create ordinary memory whereas null particles create null memory; and that ordinary memory due to massive particles with large velocities can mimic the null memory in the limit. 
Therefore, describing neutrinos through a null fluid coupled to the Einstein equations yields a good model for astrophysical sources. With D. Garfinkle in \cite{lbdg1} we introduced this model and derived the null memory effect from neutrino radiation for sources in spacetimes falling off like (\ref{safg}). As a consequence, the stress-energy of these sources enjoys a fast decay behavior. 
Whereas the latter do not yield any magnetic memory, the question remained open for sources that are falling off more slowly. 

The following investigations yield magnetic memory for Einstein-null-fluid systems describing neutrino distributions of slow decay as in 
(B-Tneutrinos) as well as (A-Tneutrinos) spacetimes. Along with that (so to say for free) comes the proof that the equations (\ref{ENF}) 
for spacetimes that are falling off at the order $O(r^{-1})$ without any assumption on the leading order decaying term, do not produce any magnetic memory.

In the following, we are going to prove for {\bf (B-Tneutrinos) spacetimes} and provide evidence for {\bf (A-Tneutrinos) spacetimes}, 
that non-isotropic neutrino distributions 
falling off at slow rates (specified in theorem \ref{T1B} and conjecture \ref{T1A}) generate  
\begin{itemize}                                                    
\item[(a)] a {\bf new magnetic memory} effect {\bf growing at rate $\sqrt{|u|}$}, sourced by the corresponding {\itshape magnetic part of the curvature} growing at the same rate. 
\item[(b)] an {\bf electric memory} effect {\bf growing at rate $\sqrt{|u|}$} sourced by the corresponding {\itshape electric part of the curvature} 
and {\itshape the $T_{\underline{L} \underline{L}}$ component of the energy-momentum tensor}, 
each growing at the same rate, and a {\itshape finite contribution from the shear}; 
\end{itemize}
Conjecture: {\bf (A-Tneutrinos) spacetimes} in addition 
\begin{itemize} 
\item[(c)] cause a {\bf new} magnetic-type memory via an integral of a curl of $T$ term, growing like $\sqrt{|u|}$. None of the memories are bounded for these spacetimes. 
\end{itemize} 
Note that the electric memory due to sources that are studied in \cite{lbdg1}, is finite. We emphasize that sources which are stationary outside a compact set do not produce any magnetic memory.

We describe the neutrinos as a null fluid in the Einstein equations (\ref{ET}), that is (\ref{ENF}), represented via its energy-momentum tensor given by 
\be \label{emnf1}
T^{\mu \nu}  = \mathcal{N} K^{\mu} K^{\nu}
\ee
with $K$ being a null vector and $\mathcal{N} = \mathcal{N}(\theta_1, \theta_2, r,  \tau_-)$ a positive scalar function depending on $r$, $\tau_-$, and the spherical variables $\theta_1, \theta_2$. 
When coupled to the Einstein equations in the most general settings, as discussed in the previous sections, the energy-momentum tensor $T^{\mu \nu}$ obeys those loose decay laws. We do not impose any symmetry nor other restrictions. 
Thus, we do not have stationarity outside a compact set, but instead a distribution of neutrinos decaying very slowly towards infinity. This distribution is very non-homogeneous and non-isotropic. 
In particular, the function $\mathcal{N}$ approaches the following structure towards spacelike infinity, that is for $r \to \infty$: 
\be \label{Auen}
\mathcal{N} = O( r^{-2} \tau^{- \frac{1}{2}} )
\ee
This follows in a straightforward manner by using (\ref{AufdemFels}) with $T^{LL}$, and the fact that $div L = tr \chi + l.o.t.$ 

Comparing the situation investigated here to the case of \cite{lbdg1}, here we find that 
the neutrino flow is more general, but that it obeys a slow convergence to the dominating behavior of the null part $T^{LL}$.

Contract the contravariant tensor $T^{\mu \nu}$ with the metric to obtain the covariant tensor $T_{\mu \nu}$. Thus it is 
$T^{LL} = \frac{1}{4} T_{\underline{L} \underline{L}}$.

\begin{The} \label{T1B}
Consider {\bf (B-Tneutrinos) spacetimes} solving (\ref{ENF}). 
The components of the energy-momentum tensor have the following decay behavior: 
\beas
T^{LL} \ & = & \ O(r^{-2} \tau_-^{- \frac{1}{2}})  \\ 
T^{AL} \ & = & \  o(r^{- \frac{5}{2}} \tau_-^{- \frac{1}{2}} )  \\ 
T^{L \underline{L}}  \ & = & \  o(r^{-3} \tau_-^{- \frac{1}{2}}) \\ 
T^{AB} \ & = & \ o(r^{-3} \tau_-^{- \frac{1}{2}}) \\ 
T^{A \underline{L}} \ & = & \ o(r^{- \frac{7}{2}} \tau_-^{- \frac{1}{2}}) \\ 
T^{\underline{L} \underline{L}} \ & = & \ o(r^{-4} \tau_-^{- \frac{1}{2}}) 
\eeas
Moreover, it is 
\beas
D_{\underline{L}}T_{A \underline{L}} & = & o(r^{- \frac{5}{2}} \tau_-^{- \frac{1}{2}} )  \\ 
D_B T_{A \underline{L}} & = & o(r^{-3} \tau_-^{- \frac{1}{2}}) \\ 
D_L T_{A \underline{L}} & = & o(r^{- \frac{7}{2}} \tau_-^{- \frac{1}{2}}) 
\eeas
\end{The}

\newpage 

\begin{conj} \label{T1A}
Consider {\bf (A-Tneutrinos) spacetimes} solving (\ref{ENF}). 
The components of the energy-momentum tensor have the following decay behavior: 
\beas
T^{LL} \ & = & \ O(r^{-2} \tau_-^{- \frac{1}{2}})  \\ 
T^{AL} \ & = & \  O(r^{- \frac{5}{2}} \tau_-^{- \frac{1}{2}} )     \\ 
T^{L \underline{L}}  \ & = & \  O(r^{-3} \tau_-^{- \frac{1}{2}}) \\ 
T^{AB} \ & = & \ O(r^{-3} \tau_-^{- \frac{1}{2}}) \\ 
T^{A \underline{L}} \ & = & \ O(r^{- \frac{7}{2}} \tau_-^{- \frac{1}{2}}) \\ 
T^{\underline{L} \underline{L}} \ & = & \ O(r^{-4}\tau_-^{- \frac{1}{2}}) 
\eeas
Moreover, it is 
\beas
D_{\underline{L}} T_{A \underline{L}} & = & O(r^{- \frac{5}{2}} \tau_-^{- \frac{1}{2}} )  \\ 
D_B T_{A \underline{L}} & = & O(r^{-3} \tau_-^{- \frac{1}{2}}) \\ 
D_L T_{A \underline{L}} & = & O(r^{- \frac{7}{2}} \tau_-^{- \frac{1}{2}}) 
\eeas
\end{conj}
It will be crucial that in (A-Tneutrinos) spacetimes $D_B T_{A \underline{L}}$ features a term of order $O(r^{-3} \tau_-^{- \frac{1}{2}})$, whereas in (B-Tneutrinos) spacetimes this is $o(r^{-3} \tau_-^{- \frac{1}{2}})$. 

{\bf Proof of Theorem \ref{T1B} and Supporting Evidence for Conjecture \ref{T1A}:} 
The decay behavior of the components of the energy-momentum tensor follows from the physical model together with the mathematical implications of the Einstein equations for the very general spacetimes.

We have 
\beas
\nabla_L e_A \ &  = & \  - \zeta_A L  \\ 
\nabla_L \underline{L} \ &  = & \  - 2 \zeta_A e_A
\eeas
From the decay behavior of $\zeta$ (given above) in the relevant spacetimes we deduce the decay along $L$ of the vectorfields $e_A$, respectively 
$\underline{L}$.

The behavior of the components of the energy-momentum tensor are a consequence of (\ref{divk1}) and (\ref{ENF}) together with (\ref{Auen}) as well as the specific decay properties of the vectorfields $e_A$ and $\underline{L}$ along $L$.

\subsection{Bianchi Equations and Structure Equations}

In this section, we give the equations for our discussion of the Einstein-null-fluid system.

The second fundamental form $k$ satisfies the equations 
\bea
tr k & = & 0   \label{k1}  \\
(curl \ k)_{ij} & = & H(W)_{ij} + \frac{1}{2} \epsilon_{ij}^{\ \ l} R_{0l}  \label{curlk1}  \\ 
(div \ k )_i & = & R_{0i}  \label{divk1}  \ . 
\eea
The null Codazzi and conjugate null Codazzi equations are 
\bea
\dlap \hat{\chi} \ & = & \  - \hat{\chi} \cdot \zeta + \frac{1}{2} (\nlap tr \chi + \zeta tr \chi) - \beta +  4\pi T_{A L} \label{codazziT1} \\ 
\dlap \hat{\underline{\chi}} \ & = & \ \hat{\underline{\chi}} \cdot \zeta + \frac{1}{2} (\nlap tr \underline{\chi} - \zeta tr \underline{\chi}) + \underline{\beta} + 
4 \pi T_{A \Lu} \ \ .   \label{codazziT2}
\eea
The propagation equations for $tr \chi$ and $tr \underline{\chi}$ along the outgoing null hypersurface $C_u$ read  
\begin{eqnarray}
\frac{dtr\underline{\chi }}{ds}  &=& - \frac{1}{2}tr\chi tr\underline{\chi }  - 2
\underline{\mu }+2\left \vert \zeta \right \vert ^{2} \label{nullstruct1} \\
\frac{dtr\chi }{ds} &=& - \frac{1}{2}\left( tr\chi \right) ^{2}   -\left \vert 
\widehat{\chi }\right \vert ^{2} -  
4 \pi T_{LL} .   \label{nullstruct2}
\end{eqnarray}
The Gauss equation for the Einstein-null-fluid case takes the form 
\begin{equation} \label{GaussENF}
K=-\frac{1}{4}tr\chi tr\underline{\chi }+\frac{1}{2}\widehat{\chi }\cdot 
\underline{\widehat{\chi }}-\rho \left( W\right) - 2 \pi T_{L \Lu}. 
\end{equation}%
Using the Gauss equation (\ref{GaussENF}), the mass aspect function $\mu$ from 
(\ref{mu}) and the conjugate mass aspect function $\underline{\mu}$ from (\ref{umu}) become 
\bea
\mu \ & = \ &  - \dlap \zeta + \frac{1}{2} \hat{\chi} \cdot \underline{\hat{\chi}} - \rho (W) - 2 \pi T_{L \Lu} \\ 
\underline{\mu} \ & = \ &   \dlap \zeta + \frac{1}{2} \hat{\chi} \cdot \underline{\hat{\chi}} - \rho (W) - 2 \pi T_{L \Lu} . 
\eea 
The propagation equation along $C_u$ for $\hat{\eta}$ is 
\beas
 \Dlap_4 \hat{\eta} + \frac{1}{2} tr \chi \hat{\eta} -  \pi T_{\Lu \Lu} \delta_{AB} &  = & 
\frac{1}{2} \alpha(W) +  \pi T_{LL} \delta_{AB} \\ 
& & 
+ \frac{1}{2} \nlap \hat{\otimes} \epsilon - \phi^{-1} \nlap^2 \phi + (a^{-1} \nlap a) \hat{\otimes} \epsilon 
+ \epsilon \hat{\otimes} \epsilon \\ 
& & 
+ \frac{3}{2} \delta \hat{\theta} - \delta \hat{\eta} 
- (\zeta -  \phi^{-1} \nlap \phi ) \hat{\otimes} \epsilon . 
\eeas
The propagation equations for $ \hat{\chi}$ and $\hat{\eta}$ in the $N$ direction within $H_t$ read 
\bea
\nlap_N \hat{\chi} & = & - \frac{1}{2} tr \chi \hat{\eta} 
- (\frac{1}{2} \delta + tr \chi) \hat{\chi} 
- \zeta \hat{\otimes} \zeta 
- \frac{1}{2} \nlap \hat{\otimes} \zeta 
- \frac{1}{2} \alpha(W) 
-  \pi \gamma_{AB} T_{LL}  \label{Nhchi1} \\ 
\nlap_N \hat{\eta} & = & -  \frac{1}{2} tr \theta \hat{\eta} 
+ (2 \delta - \frac{1}{2} tr \chi) \hat{\eta} 
+ \frac{3}{2} \delta \hat{\chi}  
+ \epsilon \hat{\otimes} \zeta -  \epsilon \hat{\otimes} \epsilon 
+ \frac{1}{2} \nlap \hat{\otimes} \epsilon 
+ \frac{1}{4} \alpha (W) 
- \frac{1}{4} \underline{\alpha} (W) . \nonumber \\ 
\label{Nheta1} 
\eea
The Bianchi equations for $\Dlap_3  \rho$ as well as $\Dlap_3 \sigma$ are 
\bea
\Dlap_3  \rho \ + \ \frac{3}{2} tr \underline{\chi} \rho \ & = & \ 
 - \dlap \underline{\beta} 
 - \frac{1}{2} \hat{\chi} \underline{\alpha} \ + \ 
 ( \epsilon  - \zeta) \underline{\beta} \ + \ 
2 \underline{\xi} \beta 
 \label{TBianchiturho3}  \\ 
 & & + \frac{1}{4} ( D_3 R_{34} - D_4 R_{33} ) \nonumber \\ 
\nonumber \\ 
\Dlap_3 \sigma \ + \ \frac{3}{2} tr \underline{\chi} \sigma \ & = & \ 
 - \clap \ \ \underline{\beta}  - \frac{1}{2} \hat{\chi} ^*\underline{\alpha}  + 
\epsilon ^*\underline{\beta} - 2 \zeta ^*\underline{\beta} 
 - 2 \underline{\xi} ^*\beta  
  \label{TBianchitusigma3}  \\ 
   & & + \frac{1}{4} ( D_{\mu} R_{3 \nu} - D_{\nu} R_{3 \mu} ) \epsilon^{\mu \nu}_{ \ \ \ 34} \nonumber 
\eea

\subsection{New Memory Effects From These Sources}

\subsubsection{Behavior along $C_u$ and Limits at $\mathcal{I}^+$}

\begin{The} \label{conclcurvandfieldcompts1} 
For {\bf (B-Tneutrinos) spacetimes}, the normalized curvature components $r\underline{\alpha }\left( W\right) $, $r^{2}\underline{\beta }\left(
W\right) $, and normalized energy-momentum component
$r^2 T_{33}$ 
have limits on $C_u$ as $t\rightarrow \infty $. \\
They are 
\begin{eqnarray*}
\lim_{C_{u},t\rightarrow \infty }r\underline{\alpha }\left( W\right)
&=&A_{W}\left( u,\cdot \right) ,\, \ \ \ \ \ \ \ \ \ \ \ \
\lim_{C_{u},t\rightarrow \infty }\,r^{2}\underline{\beta }\left( W\right)
=\underline{B}_{W}\left( u,\cdot \right) \\
\lim_{C_{u},t\rightarrow \infty }r^2 T_{33} 
&=& \mathcal{T}_{33} \left( u,\cdot \right) , 
\end{eqnarray*}
where the limits are on $S^{2}$ and depend on $u$. Moreover, these limits satisfy 
\begin{eqnarray*}
\left| A_{W}\left( u,\cdot \right) \right| &\leq &C\left( 1+\left| u\right|
\right) ^{-3/2}\, \, \ \ \ \ \ \ \ \ \ \ \ \ \left| \underline{B}_{W}\left( u,\cdot
\right) \right| \leq C\left( 1+\left| u\right| \right) ^{-1/2} \\
\mathcal{T}_{33} \left( u,\cdot \right)  &\leq &C\left( 1+\left| u\right|
\right) ^{- \frac{1}{2}}   . \ \
\end{eqnarray*}
\end{The} 
These are direct consequences from the results in \cite{lydia1, lydia2} but allowing for large data (see introduction of the present article) together with theorem \ref{T1B} above for theorem \ref{conclcurvandfieldcompts1} and conjecture \ref{T1A} from above for conjecture \ref{conclcurvandfieldcompts1conj2}. 
\begin{conj} \label{conclcurvandfieldcompts1conj2}
For {\bf (A-Tneutrinos) spacetimes}, the normalized curvature components $r\underline{\alpha }\left( W\right) $, $r^{2}\underline{\beta }\left(
W\right) $, and normalized energy-momentum components 
$r^2 T_{33}, r^3 T_{43}$, 
$r^3 T_{AB}, r^4 T_{44}$ 
as well as the derivatives 
$D_A T_{3B}$ 
have limits on $C_u$ as $t\rightarrow \infty $. \\
They are 
\begin{eqnarray*}
\lim_{C_{u},t\rightarrow \infty }r\underline{\alpha }\left( W\right)
&=&A_{W}\left( u,\cdot \right) ,\, \ \ \ \ \ \ \ \ \ \ \ \
\lim_{C_{u},t\rightarrow \infty }\,r^{2}\underline{\beta }\left( W\right)
=\underline{B}_{W}\left( u,\cdot \right) \\
\lim_{C_{u},t\rightarrow \infty }r^2 T_{33} 
&=& \mathcal{T}_{33} \left( u,\cdot \right) , 
\end{eqnarray*}
and 
\begin{eqnarray*}
\lim_{C_{u},t\rightarrow \infty }r^3 T_{43} 
&=& \mathcal{T}_{43} \left( u,\cdot \right) , \\
\lim_{C_{u},t\rightarrow \infty } r^3 T_{AB} 
&=& \mathcal{T}_{AB}
 \left(
u,\cdot \right) ,\  \\ 
\lim_{C_{u},t\rightarrow \infty }r^4 T_{44} 
&=& \mathcal{T}_{44} \left( u,\cdot \right) , \\
\lim_{C_{u},t\rightarrow \infty }r^3 \nlap_A T_{3B} 
&=& (\nlap_A T_{3B})^*  \left( u,\cdot \right) , 
\end{eqnarray*}
where the limits are on $S^{2}$ and depend on $u$. Moreover, these limits satisfy 
\begin{eqnarray*}
\left| A_{W}\left( u,\cdot \right) \right| &\leq &C\left( 1+\left| u\right|
\right) ^{-3/2}\, \, \ \ \ \ \ \ \ \ \ \ \ \ \left| \underline{B}_{W}\left( u,\cdot
\right) \right| \leq C\left( 1+\left| u\right| \right) ^{-1/2} \\
\mathcal{T}_{33} \left( u,\cdot \right)  &\leq &C\left( 1+\left| u\right|
\right) ^{- \frac{1}{2}} \\ 
(\nlap_A T_{3B})^*  \left( u,\cdot \right) 
 &\leq &C\left( 1+\left| u\right|
\right) ^{- \frac{1}{2}} 
  . \ \
\end{eqnarray*}
\end{conj}

\begin{The} \label{conclXi1}
For (B-Tneutrinos) spacetimes, on each null hypersurface $C_u$, the limit of $r\widehat{{\underline{\chi }}  }$ exists as $%
t\rightarrow \infty $, in particular 
\begin{equation*}
- \frac{1}{2} \lim_{C_{u},t\rightarrow \infty }r\widehat{\underline{\chi }} = 
\Xi \left( u,\cdot \right) 
\end{equation*}
with limit $\Xi $ being a symmetric traceless 2-covariant tensor on $S^{2}$ 
depending on $u$ and satisfying 
\begin{equation*}
\left| \Xi \left( u,\cdot \right) \right| _{\overset{\circ }{\gamma }}\leq
C\left( 1+\left| u\right| \right) ^{-1/2}.
\end{equation*}
Moreover, we have 
\begin{eqnarray}
\frac{\partial \Xi }{\partial u} &=&-\frac{1}{4}A_{W}  \label{XiAT1} \\ 
Chi_3 &=& - \Xi  \label{Sigmau*1} \label{XiChi3T1} \\ 
\underline{B} &=& - 2 \dlap \Xi \label{XiBT1} 
\end{eqnarray}
\end{The}
{\bf Proof:} 
Concerning the latter limits: 
(\ref{XiBT1}) follows from (\ref{codazziT2})
as $T_{A \underline{L}}$ decays faster along $C_u$ than the leading order terms. 
(\ref{XiChi3T1}) follows from (\ref{Nhchi1}), and 
(\ref{XiAT1}) from (\ref{Nheta1}). Hereby, we use theorem \ref{T1B} for (B-Tneutrinos) spacetimes. \\ 

In a similar way, the following conjecture follows using statements \ref{T1A} for (A-Tneutrinos) spacetimes.

\begin{conj} \label{conj2conclXi1}
For (A-Tneutrinos) spacetimes, the corresponding statements of theorem \ref{conclXi1} hold. 
\end{conj}

\begin{The}
(A-Tneutrinos) spacetimes. Consider spacetimes with initial data where $\bar{g} - \eta = O(r^{- \frac{1}{2}})$. In this case, if the solution is well-behaved, the following holds. 
At future null infinity $\mathcal{I}^+$, it is 
\bea 
- 2 \clap \ \ \dlap \Xi \ & = & \  \clap \ \ \underline{B} \ + \ 4 \pi \big{(} \clap \ \ T  \big{)}^*_{34_3}  \label{curldivXi1} \\ 
- 2 \dlap \dlap \Xi \ & = & \ \dlap \underline{B} \label{divdivXi1}
\eea

\end{The}

{\bf \itshape Proof:} 
To prove this theorem, we consider equation (\ref{codazziT2}) for $\underline{\hat{\chi}}$ 
\[
\dlap \hat{\underline{\chi}} \  =  \ \hat{\underline{\chi}} \cdot \zeta + \frac{1}{2} (\nlap tr \underline{\chi} - \zeta tr \underline{\chi}) + \underline{\beta} + 
4 \pi T_{A \Lu} 
\]
Take the $\clap \ \ $ of this equation, multiply with $r^3$ and take the limit on $C_u$ as $t \to \infty$ to obtain equation (\ref{curldivXi1}). 
Similarly, take the $\dlap$ of (\ref{codazziT2}) multiply with $r^3$ and take the limit on $C_u$ as $t \to \infty$, which yields equation (\ref{divdivXi1}). \\ 

{\bf Remark:} The fact that the energy-momentum tensor component $T_{A \underline{L}}$ in (\ref{codazziT2}) produces a curl contribution in (\ref{curldivXi1}) is unique to the spacetimes with a metric decaying like $O(r^{- \frac{1}{2}})$ towards infinity, thus for (A-Tneutrinos) spacetimes. If we assume just a little more decay such as $o(r^{- \frac{1}{2}})$, as for (B-Tneutrinos) spacetimes, then the curl of $T$ decays faster and the limiting equation (\ref{curldivXi1}) reduces to 
\be
- 2 \clap \ \ \dlap \Xi \  =  \  \clap \ \ \underline{B}  \label{curldivXi2}
\ee
Note that the divergence on $S$ of the same component $T_{A \underline{L}}$ is of lower order and therefore there is no $T$-term in (\ref{divdivXi1}). 
In fact, equations (\ref{divdivXi1}) and (\ref{curldivXi2}) have been known to hold for sources that are stationary outside a compact set. 
We point out that they do hold as well for the more general decay as in (B-Tneutrinos). However, for the most general class of spacetimes (A-Tneutrinos) with a metric decaying like $O(r^{- \frac{1}{2}})$ towards infinity, the curl contribution of $T$ kicks in.

\subsubsection{Growing Electric Memory}
\label{TEM}

Next, we are going to derive the detailed structures of the electric memory in (B-Tneutrinos) as well as (A-Tneutrinos) spacetimes. The only difference in radiation between the two is that in the former case the null memory from the shear is finite, but in the latter case unbounded. The ensuing investigations prove theorem \ref{theABT1} and provide evidence for conjecture \ref{conjtheABT1}. 

Consider the Bianchi equation (\ref{TBianchiturho3}) for the electric component $\rho$ of the curvature. Focussing on the highest order terms, we write 
\bea
\Dlap_3  \rho \ + \ \frac{3}{2} tr \underline{\chi} \rho \ & = & \ 
 - \dlap \underline{\beta} 
 - \frac{1}{2} \hat{\chi} \underline{\alpha} 
 - 2 \pi  D_4 T_{33}  \  + \ l.o.t. \label{Bianchiturho66} 
\eea
We take into account that 
\beas
- \frac{1}{4} D_4 R_{33}  \ & = & \  - 2 \pi  D_4 T_{33} 
\eeas 
and the leading order term in the latter is given by 
\beas
  - 2 \pi (\Dlap_4 \mathcal{N}) 
 \ & = & \  + 2 \pi tr \chi \mathcal{N}  \ . 
 \eeas 
Therefore we have 
\beas
  - 2 \pi  D_4 T_{33} 
  \ & = & \ + 2 \pi tr \chi T_{33} + l.o.t.  
\eeas 
Thus we write (\ref{Bianchiturho66}) as 
\bea
\rho_3  + \frac{\partial}{\partial u} (\hat{\chi} \cdot \hat{\underline{\chi}})  \  & = &   \ 
 - \dlap \underline{\beta} + \frac{1}{4} tr \chi  |\hat{\underline{\chi}}|^2  + 2 \pi tr \chi T_{33} + l.o.t. =   O(r^{-3} \tau_-^{- \frac{1}{2}}) 
   \label{Bianchiturho77} 
\eea 
Taking the limit of (\ref{Bianchiturho77}) after multiplying the equation with $r^3$ yields 
\be \label{LrhoT1*}
\mathcal{P}_3  \ = \ 
- \dlap \underline{B} + 2 | \Xi |^2 + 4 \pi \mathcal{T}_{33} 
\ee
Now, similar to the situation investigated in section \ref{FNI}, we integrate equation (\ref{LrhoT1*}) with respect to $u$, using (\ref{divdivXi1}), to obtain 
\be \label{supergoldT1}
\dlap \dlap (Chi^- - Chi^+)   \ = \ 
(\mathcal{P}^- - \mathcal{P}^+) - \int_{- \infty}^{+ \infty} \big{(} | \Xi |^2  \ + \  2 \pi  \ \mathcal{T}_{33} \big{)} \ du 
\ee
Recall (\ref{FXiT1}). 
In particular, there exists a function $\Phi$ on $S^2$ such that 
\bea
 \dlap (Chi^- - Chi^+) & = & \nlap \Phi  + X \label{supergoldT1**} \\ 
\dlap \dlap (Chi^- - Chi^+) & = & \slap \Phi  \nonumber \\ 
& = & (\mathcal{P} - \bar{\mathcal{P}})^- - (\mathcal{P} - \bar{\mathcal{P}})^+ \nonumber \\ 
& & - 2 (F_T - \bar{F}_T)  \label{supergoldT3**} 
\eea
with $X = \nlap^{\perp} \Psi$ for a function $\Psi$ whose Laplacian $\slap \Psi = \clap \ \  \dlap (Chi^- - Chi^+)$. 
Similarly as in section \ref{FNI}, in order to reveal the new structures more clearly, we first put the contribution from $\Psi$ to zero. In section \ref{TMM}, we treat the most general case with $\nlap^{\perp} \Psi \neq 0$ to obtain the complete general result and the full set of equations 
(\ref{TPsi1})-(\ref{TPhi33}). In equations (\ref{TPsi1})-(\ref{TPsi2}) in the system (\ref{TPsi1})-(\ref{TPhi33}) the magnetic memory occurs naturally with corresponding contributions from the stress-energy tensor. 

We find the same structures for $\mathcal{P}$ as in (\ref{limitstructures2}). 
In addition to the behavior already found and described in section \ref{FNI} for the Einstein vacuum equations, we deduce from (\ref{supergoldT1}) and theorem \ref{conclcurvandfieldcompts1} that the null memory due to the integral of the null limit $\mathcal{T}_{33}$ of the neutrino distribution grows like $\sqrt{|u|}$. This behavior is very different from our results in \cite{lbdg1} as in \cite{lbdg1} the corresponding contribution from neutrino radiation is finite. 

We have {\bf proven} the following theorem. 

\begin{The} \label{theABT1}
The following holds for (B-Tneutrinos) spacetimes. 

If $\nlap^{\perp} \Psi = 0$, then  
$(Chi^- - Chi^+)$ is determined by equation (\ref{supergoldT1**}) on $S^2$ where $\Phi$ is the solution with vanishing mean of (\ref{supergoldT3**}). 

\end{The}

\begin{conj} \label{conjtheABT1}
The statement of theorem \ref{theABT1} holds correspondingly for (A-Tneutrinos) spacetimes. 
\end{conj}

\subsubsection{Rotation: Growing Magnetic Memory}
\label{TMM}

Next, we investigate the most general case with $\nlap^{\perp} \Psi \neq 0$. 
In this section, we establish the magnetic memory. Thereby, the ultimate class of solutions exhibits extra structures generating a qualitatively completley new type of this memory. 
We investigate the following for (A-Tneutrinos) spacetimes. The (B-Tneutrinos) spacetimes do not feature the memory sourced by the curl of $T$, but they do feature all the other memory components. 
The following arguments prove theorem \ref{theneutrinoscurl55} and provide evidence for conjecture \ref{conjneutrinoscurl55}. 

Consider the Bianchi equation (\ref{TBianchitusigma3}) for the magnetic component $\sigma$ of the curvature. Focussing on the highest order terms, we have 
\bea
\Dlap_3 \sigma \ + \ \frac{3}{2} tr \underline{\chi} \sigma \ & = & \ 
 - \clap \ \ \underline{\beta}  - \frac{1}{2} \hat{\chi} ^*\underline{\alpha}  
 + 4 \pi 
(\clap \ \ T)_{34_3}   \  + \ l.o.t. 
  \label{Bianchitusigma66}
\eea
A short computation yields 
\bea
\sigma_3  + \frac{\partial}{\partial u} ( \hat{\chi} \wedge \hat{\underline{\chi}} ) \ = \ - \clap \ \ \underline{\beta} 
 + 4 \pi 
(\clap \ \ T)_{34_3}   \  + \ l.o.t. 
\ = \ 
O(r^{-3} \tau_-^{- \frac{1}{2}}) \label{Bianchitusigma77}
\eea
In the following, the null limit of $r^3 (\clap \ \ T)_{34_3}$ on $C_u$ as $t \to \infty$ is denoted as 
\[ 
(\clap \ \ T)^*_{34_3} = ( (\nlap_A T_{3B})^* - (\nlap_B T_{3A})^* ) \ . 
\]
Next, multiply equation (\ref{Bianchitusigma77}) by $r^3$ and take the limit on $C_u$ as $t \to \infty$ to get  
\[
\mathcal{Q}_3  \ = \ 
- \clap  \ \ \underline{B} 
+ 4 \pi \big{(} \clap \ \ T  \big{)}^*_{34_3} 
\] 
Using (\ref{curldivXi1}) 
gives 
\be \label{LsigmaT1*}
\mathcal{Q}_3  \ = \ 
2 \ \clap \ \ \dlap  \Xi \ 
+ \ 8 \pi \big{(} \clap \ \ T  \big{)}^*_{34_3} 
\ee 
Similar to the situation investigated in section \ref{new}, we integrate equation (\ref{LsigmaT1*}) with respect to $u$ to obtain 
\be \label{2supergoldsigmaT1}
\clap \ \  \dlap (Chi^- - Chi^+) \ = \ 
(\mathcal{Q}^- - \mathcal{Q}^+) 
\ + \ 
4 \pi \int_{- \infty}^{+ \infty} \big{(} \clap \ \ T  \big{)}^*_{34_3}  \ du 
\ee
We find the structures of $\mathcal{Q}$ to be as in (\ref{limitstructures2dawn}). 
In (\ref{2supergoldsigmaT1}) we show that there is a new contribution to the magnetic null memory due to the integral of the null limit $\big{(} \clap \ \ T  \big{)}^*_{34_3}$ of the general neutrino distribution. The last term in (\ref{2supergoldsigmaT1}) is infinite. More precisely, fix $u_0$, then the integral 
$\int_{u_0}^{u} \big{(} \clap \ \ T  \big{)}^*_{34_3}  \ du$ 
diverges like $\sqrt{|u|}$ as $|u| \to \infty$. This is very different to the situation we derived in \cite{lbdg1}, where there is no magnetic memory, because the relevant components of the energy-momentum tensor decay too fast to produce a limit. In \cite{lbdg1}, all the memory is of electric parity only, there is no magnetic memory. 

Equation (\ref{2supergoldsigmaT1}) shows that the new magnetic memory is due to $(\mathcal{Q}^- - \mathcal{Q}^+)$ and $\int_{- \infty}^{+ \infty} \big{(} \clap \ \ T  \big{)}^*_{34_3}  \ du$. 

Recall the quantities (\ref{FXiT1}) and (\ref{RcurlT1}) introduced in section \ref{setting}.

Now, the new equations for neutrino sources (\ref{supergoldT1}) and (\ref{2supergoldsigmaT1}) yield the system 
\bea
 \dlap (Chi^- - Chi^+) & = & \nlap \Phi + \nlap^{\perp} \Psi  \label{TPsi1} \\ 
\clap \ \  \dlap (Chi^- - Chi^+) & = & \slap \Psi  \nonumber \\ 
& = & (\mathcal{Q} - \bar{\mathcal{Q}})^- - (\mathcal{Q} - \bar{\mathcal{Q}})^+ \nonumber \\ 
& & + (\mathcal{A}_T - \bar{\mathcal{A}}_T)     \label{TPsi2} \\ 
\dlap \dlap (Chi^- - Chi^+) & = & \slap \Phi  \nonumber  \\ 
& = & (\mathcal{P} - \bar{\mathcal{P}})^- - (\mathcal{P} - \bar{\mathcal{P}})^+ \nonumber \\ 
& & - 2 (F_T - \bar{F}_T)  \label{TPhi33} 
\eea
with the structures (\ref{limitstructures2}) for $\mathcal{P}$ and (\ref{limitstructures2dawn}) for $\mathcal{Q}$. 
This system is solved by Hodge theory on the sphere $S^2$. 

Equations (\ref{TPsi1})-(\ref{TPhi33}) describe the full picture of gravitational wave memory in the general spacetimes. In addition to the effects for the Einstein vacuum equations (\ref{EV}), that is the electric memory sourced by $\mathcal{P}$ and $F$, as well as the magnetic memory sourced by $\mathcal{Q}$; for the equations (\ref{ENF}) describing general neutrino distributions, there is a growing contribution from the integral of $\mathcal{T}_{33}$ to the electric part and a growing contribution from the integral of $\big{(} \clap \ \ T  \big{)}^*_{34_3}$ to the magnetic memory. 

These {\bf prove} the following theorem \ref{theneutrinoscurl55} and provide evidence for conjecture \ref{conjneutrinoscurl55}.

\begin{The} \label{theneutrinoscurl55}
In (B-Tneutrinos) spacetimes, 
$(Chi^- - Chi^+)$ is determined by equation (\ref{TPsi1}) on $S^2$ where $\Phi$ solves (\ref{TPhi33}) 
and $\Psi$ solves (\ref{TPsi2}) but with $\mathcal{A}_T$ as well as $\bar{\mathcal{A}}_T$ being identically zero. 
\end{The}

\begin{conj} \label{conjneutrinoscurl55}
In (A-Tneutrinos) spacetimes, 
$(Chi^- - Chi^+)$ is determined by equation (\ref{TPsi1}) on $S^2$ where $\Phi$ solves (\ref{TPhi33}) 
and $\Psi$ solves (\ref{TPsi2}). 
\end{conj}

\section{Conclusions}
\label{conclusions}

We have derived new structures in gravitational waves and gravitational wave memories. Thereby, new memory effects of various types have emerged. 

The most enthralling new property is the growing magnetic memory. 
Such an effect does not exist (not even in finite form) in systems with a decay like $O(r^{-1})$. 

(Bvac) and (Avac) spacetimes exhibit a magnetic memory growing with $\sqrt{|u|}$. Thus, it is interesting to point out that magnetic memory arises naturally in these spacetimes in pure gravity. 
The electric memory in these spacetimes diverges as well at the same rate. The null part of the electric memory is due to shear being finite for (Bvac) and unbounded for (Avac). A good source for the new phenomena is provided by the non-isotropic dynamics of neutrinos, that are not stationary outside a compact set but whose distribution decays very slowly towards infinity. We describe these neutrino systems by coupling the Einstein equations to a null fluid, yielding spacetimes of the types  (B-Tneutrinos) as well as (A-Tneutrinos). Investigating these, we prove that the aforementioned memories exist also in the Einstein-null-fluid system for neutrinos. In addition, the neutrinos cause two new, growing effects. One is the electric null memory sourced by the $T_{33}$ component of the energy-momentum tensor diverging at the highest rate. This is in contrast with systems that fall off like $O(r^{-1})$, as in \cite{lbdg1}, where the contribution is finite. The other effect from neutrinos is of completely different nature and new even from a qualitative point of view, it 
appears for (A-Tneutrinos) spacetimes: the magnetic memory is powered by a new component sourced by the curl of $T$ and diverging at the highest rate. 

Besides the leading effects, we show that there is a panorama of finer structures. 
First, we see that 
the leading order, growing components as well as all the remaining growing components in the limits $\mathcal{P}$ and $\mathcal{Q}$ originate from the corresponding curvature parts. On top of that, they also produce finite memories. Second, the components of $\mathcal{P}$ and $\mathcal{Q}$ sourced by $\hat{\chi} \cdot \hat{\underline{\chi}}$, respectively $\hat{\chi} \wedge \hat{\underline{\chi}}$, generate finite memories too. We re-emphasize that these properties are unique to spacetimes with slow fall-off.

To study physical situations of matter distributions, we require the no incoming radiation condition at past null infinity $\mathcal{I}^-$. Then we let the initial data evolve under the coupled Einstein-matter equations. Outgoing radiation of these systems produces the memories at future null infinity $\mathcal{I}^+$. We also find that data as in \cite{lydia1, lydia2} extended towards the past, produces the corresponding effects at past null infinity $\mathcal{I}^-$. Assuming the no incoming radiation condition switches off the latter. 

Our new gravitational wave memories can be used to detect, identify and gain more information about sources. In principle, they should be seen in present and future gravitational wave detectors. 

The new results bring many applications. 
If dark matter behaves like neutrinos or similar matter that is non-isotropic and non-stationary outside a compact set, but decays very slowly, then the new phenomena 
can be used to detect dark matter via gravitational waves. Certain applications as well as the possibility to see these effects in gravitational wave detectors are discussed in \cite{lydia14}. 

The results for the (B-Tneutrinos) spacetimes and  (A-Tneutrinos) spacetimes hold for any Einstein-matter system with an energy-momentum tensor of a source that is non-isotropic and obeys specific decay laws.  

The dynamics of compact objects in General Relativity leave characteristic footprints in the universe. By investigating sources that are not stationary outside a compact set but decay slowly, a new landscape has appeared. These more general dynamics carry further structures that show in gravitational radiation.

\section*{Acknowledgments} 

The author thanks Demetrios Christodoulou for useful remarks on a draft of this article. 
The author thanks the NSF and the Simons Foundation; 
the author was supported by the NSF Grant No. DMS-1811819 
and 
the Simons Fellowship in Mathematics No. 555809.  \\ \\ \\ \\




\begin{thebibliography}{99} 
\bibitem{ligodetect1}
B. P. Abbott {\it et al.} (LIGO Scientific Collaboration and Virgo Collaboration)
, Phys. Rev. Lett. {\bf 116}, 061102 (2016) 
\bibitem{ligodetect2}
B. P. Abbott {\it et al.} (LIGO Scientific Collaboration and Virgo Collaboration)
, Phys. Rev. Lett. {\bf 116}, 241102 (2016) 
\bibitem{ligodetect3}
B. P. Abbott {\it et al.} (LIGO Scientific Collaboration and Virgo Collaboration)
, Phys. Rev. Lett. {\bf 118}, 221101 (2017) 
\bibitem{lydia1} L. Bieri.  
        \begin{itshape} An Extension of the Stability Theorem of the Minkowski Space
in General Relativity. \end{itshape}
        ETH Zurich, Ph.D. thesis.  \textbf{17178}. 
        Zurich. (2007).  
\bibitem{lydia2} L. Bieri.  
        \begin{itshape} Extensions of the Stability Theorem of the Minkowski Space
in General Relativity. Solutions of the Einstein Vacuum Equations. \end{itshape}
        AMS-IP. Studies in Advanced Mathematics. Cambridge. MA. (2009).      
\bibitem{lydia69} L. Bieri.   
 \begin{itshape} An Extension of the Stability Theorem of the Minkowski Space in General Relativity. \end{itshape} 
 Journal of Differential Geometry.  \textbf{86}.  no.1. (2010). 17-70.     
\bibitem{lydia4} L. Bieri.          
    \begin{itshape} Answering the Parity Question for Gravitational Wave Memory. \end{itshape} 
    Phys. Rev. D 98. 124038. (2018).
 \bibitem{lydia14} L. Bieri.    
 {\itshape ``New Effects in Gravitational Waves and Memory."} 
Phys. Rev. D 103. 024043. (2021).
\bibitem{1lpst1} L. Bieri, P. Chen, S.-T. Yau. 
  \begin{itshape}Null Asymptotics of Solutions of the Einstein-Maxwell Equations in
General Relativity and 
  Gravitational Radiation.\end{itshape}
  Advances in Theor. and Math.Phys.15.4. (2011). 
\bibitem{1lpst2} L. Bieri, P. Chen, S.-T. Yau. 
  \begin{itshape}  The Electromagnetic Christodoulou Memory Effect and its Application to Neutron Star Binary Mergers.   \end{itshape} 
 Class.Quantum Grav. 29, 21, (2012).  
\bibitem{lbdg1} L. Bieri, D. Garfinkle. 
\begin{itshape} Neutrino Radiation Showing a Christodoulou Memory Effect in General Relativity.   \end{itshape} 
Annales Henri Poincar\'e. 23. 14. 329. (2014). (DOI 10.1007/s00023-014-0329-1). 
\bibitem{lbdg3} L. Bieri, D. Garfinkle. 
  \begin{itshape}   Perturbative and gauge invariant treatment of gravitational wave memory.  \end{itshape} 
Phys. Rev. D. 89. 084039. (2014). 
\bibitem{lbdg2} L. Bieri, D. Garfinkle. 
\begin{itshape}  An electromagnetic analog of gravitational wave memory. \end{itshape} 
Class. Quantum Grav. 30. 19. (2013) 195009.  
\bibitem{bgsty1} L. Bieri, D. Garfinkle, S.-T. Yau. 
  \begin{itshape}   Gravitational wave memory in de Sitter spacetime.  \end{itshape} 
  Phys. Rev. D 94. no.6. (2016) 064040
\bibitem{BGYmemcosmo1} L. Bieri, D. Garfinkle, N. Yunes. 
 \begin{itshape} Gravitational wave memory in $\Lambda$CDM cosmology. \end{itshape} 
Classical and Quantum Gravity. 34. 21. (2017). 215002
\bibitem{lbatdgbw} L. Bieri, A. Tolish, D. Garfinkle, R. Wald.  
{\itshape Examination of a simple example of gravitational wave memory.} 
Phys. Rev. D 90. 044060. (2014).  
\bibitem{blda1} L. Blanchet, T. Damour. 
\begin{itshape} Postnewtonian Generation of Gravitational Waves.  \end{itshape} 
Ann.Inst. H. Poincar\'e. Theor. 50. 377. (1989). 
\bibitem{blda2} L. Blanchet, T. Damour. 
\begin{itshape} Hereditary effects in gravitational radiation. \end{itshape} 
Phys.Rev.D 46. 304. (1992). 
\bibitem{braginskyg}
V.B. Braginsky and L.P. Grishchuk, Sov. Phys. JETP, {\bf 62}, 427 (1985) 
\bibitem{braginsky} V.B. Braginsky, K.S. Thorne. 
Nature (London) {\bf 327}, 123. (1987). 
\bibitem{ychb1}  Y. Choquet-Bruhat. 
        \begin{itshape} Th\'{e}or\`{e}me d'existence pour certain syst\`{e}mes d'equations 
        aux d\'{e}riv\'{e}es partielles nonlin\'{e}aires. \end{itshape}
        Acta Math. \textbf{88}. 
        (1952). 141-225. 
        \bibitem{ychb3}  Y. Choquet-Bruhat. 
        \begin{itshape} Beginning of the Cauchy problem for Einstein's field equations.  \end{itshape}
        Surveys in Differential Geometry. \textbf{20}. International Press. (2015). 

\bibitem{ychb2} Y. Choquet-Bruhat. P.T. Chru\'sciel, J. Loizelet. 
     \begin{itshape} Global solutions of the Einstein-Maxwell equations in higher dimensions. \end{itshape}
 Class.Quant.Grav. 23. (2006). 7383-7394. 
\bibitem{bruger}  Y. Choquet-Bruhat, R. Geroch. 
        \begin{itshape} Global Aspects of the Cauchy Problem in General Relativity.  \end{itshape}
        Comm.Math.Phys. \textbf{14}. 
        (1969). 329-335. 
\bibitem{chrmemory}  D. Christodoulou. 
        \begin{itshape} Nonlinear Nature of Gravitation and Gravitational-Wave
Experiments. \end{itshape}
        Phys.Rev.Letters. \textbf{67}. 
        (1991). no.12. 1486-1489. 
\bibitem{chrIV2000}  D. Christodoulou. 
        \begin{itshape} The Global Initial Value Problem in General Relativity. \end{itshape} 
        Proceedings of the 9th Marcel Grossmann Meeting on General Relativity. Rome. Italy. (2000). 
\bibitem{DCblh2008} D. Christodoulou. 
   \begin{itshape} The formation of black holes in general relativity. \end{itshape} EMS Monographs in Mathematics. European Mathematical Society (EMS), Zurich, (2009). MR2488976 (2009k:83010)
\bibitem{sta} D. Christodoulou, S. Klainerman.
        \begin{itshape} The global nonlinear stability of the Minkowski space.
\end{itshape}
        Princeton Math.Series \textbf{41}. 
        Princeton University Press. Princeton. NJ. (1993).   
\bibitem{dam1} T. Damour. 
 \begin{itshape}  Analytical calculations of gravitational radiation.    \end{itshape}
    Proc. 4th Marcel Grossmann Meeting. Part A. (1986). 365.         
   \bibitem{favata}
M. Favata, Class. Quantum Grav. {\bf 27}, 084036 (2010)        
\bibitem{flanagan}
E. Flanagan and D. Nichols, Phys. Rev. D {\bf 92}, 084057 (2015)
\bibitem{jorg}
J. Frauendiener, Class. Quantum Grav. {\bf 9}, 1639 (1992)

\bibitem{fried1} H. Friedrich. 
        \begin{itshape} On the Existence of $n$-Geodesically Complete or Future Complete Solutions 
        of Einstein's Field Equations with Smooth Asymptotic Structure. \end{itshape}
        Comm.Math.Phys. \textbf{107}. (1986). 587-609.  

           \bibitem{vashin2017} P. Hintz, A. Vasy. 
        \begin{itshape} A global analysis proof of the stability of Minkowksi space and the polyhomogeneity of the metric. 
\end{itshape}
     Annals of PDE. 6:2 (2020).      

           
           
\bibitem{Lasky1} 
  P.~D.~Lasky, E.~Thrane, Y.~Levin, J.~Blackman and Y.~Chen,
  Phys.\ Rev.\ Lett.\  {\bf 117}, no. 6, (2016)
  
  
  
\bibitem{winma2} T. M\"adler, J. Winicour. 
  \begin{itshape} The sky pattern of the linearized gravitational memory effect. 
         \end{itshape} 
         Classical and Quantum Gravity. \textbf{33}. 17. (2016). 


\bibitem{WaldTm1} G. Satishchandran, R.M. Wald. 
(2019). 
https://arxiv.org/abs/1901.05942


\bibitem{strominger}
A. Strominger and Zhiboedov, JHEP {\bf 1601}, 086 (2016)

\bibitem{thorne}
K. Thorne, in {\it Gravitational Radiation}, eds. N. Deruelle and T. Piran (North Holland, Amsterdam, 1983)

\bibitem{thorne2}
K.S. Thorne, Phys. Rev. D {\bf 45}, 520 (1992)


\bibitem{tolwal1} A. Tolish,  R. M. Wald. 
Phys. Rev. D {\bf 89}, 064008 (2014). 
\bibitem{twcosmo}
A. Tolish and R. M. Wald, Phys. Rev. D {\bf 94}, 044009 (2016)


\bibitem{winicour}
J. Winicour, Class. Quantum Grav. {\bf 31}, 205003 (2014)

\bibitem{will} A. G. Wiseman, C. M. Will. 
Phys. Rev. D {\bf 44}, R2945 (1991).


\bibitem{zeldovich}
Ya.B. Zel'dovich and A.G. Polnarev, Sov. Astron. {\bf 18}, 17 (1974)
\bibitem{zip} N. Zipser. 
        \begin{itshape} The Global Nonlinear Stability of the Trivial Solution of
the Einstein-Maxwell Equations.  \end{itshape}
        Ph.D. thesis. Harvard Univ. Cambridge MA. (2000).          
\bibitem{zip2} N. Zipser.  
        \begin{itshape} Extensions of the Stability Theorem of the Minkowski Space
in General Relativity. - Solutions of the Einstein-Maxwell Equations.
\end{itshape}
        AMS-IP. Studies in Advanced Mathematics. Cambridge. MA. (2009).          
\end{thebibliography}
\end{document}